# OXYGEN ABUNDANCES IN
# LOW SURFACE BRIGHTNESS DISK GALAXIES

Stacy S. McGaugh[1,2]

Department of Astronomy, University of Michigan, Ann Arbor, MI 48109

[2]Current address: Institute of Astronomy, University of Cambridge, Madingley Road, Cambridge CB3 0HA, England. I: ssm@mail.ast.cam.ac.uk

# ABSTRACT

The oxygen abundances in the H II regions of a sample of low surface brightness (LSB) disk galaxies are presented. In general, LSB galaxies are found to be metal poor ($Z < \frac{1}{3}Z_\odot$). Indeed, some LSB galaxies rival the lowest abundance extragalactic objects known. These low metallicities indicate that LSB galaxies evolve slowly, forming relatively few stars over a Hubble time.

The low metallicities of LSB galaxies occur even though many are comparable in size and mass to the prominent spirals which define the Hubble sequence. As well as being low in surface brightness, these galaxies tend to be isolated. This suggests that surface mass density and environment are more relevant to galaxy evolution than gross size.

Despite the low surface brightness of the disks, massive ($M > 60 M_\odot$) stars are inferred to be present and no abnormality of the IMF is indicated. Many low excitation H II regions exist at low metallicity in LSB galaxies, and the ionization parameter is not tightly correlated with metallicity. However, there does seem to be a significant envelope of maximum ionization at a given metallicity.

*Subject headings:* galaxies: abundances — galaxies: evolution — nebulae: abundances — nebulae: H II regions

# 1. INTRODUCTION

Low Surface Brightness galaxies are an important but often neglected part of the galaxy content of the universe. Their importance stems both from the selection effects which cause them to be under-represented in galaxy catalogs, and from the clues they contain about the physical processes of galaxy evolution. Here, emphasis is placed on the latter, as probed by the gas phase chemical abundances in LSB galaxies. Though the study of abundances in high surface brightness (HSB) spirals is a well developed field (e.g., Pagel & Edmunds 1981; McCall, Rybski, & Shields 1985; Torres-Peimbert, Peimbert, & Fierro 1989; Diaz et al. 1991; Vila-Costas & Edmunds 1992), at present only a very little is known about the composition of a limited sample of LSB spirals (Webster et al. 1983).

That LSB galaxies have been neglected is not surprising considering the technical difficulties imposed by their nature. With *peak* surface brightnesses $\mu_0 \geq 23$ $B$ mag arcsec$^{-2}$, spectral observations of their stellar continua are virtually impossible because the sky background always dominates the signal. This problem is approached here by concentrating on the emission regions within LSB galaxies for which it is possible, though still difficult, to obtain reasonable quality data.

To this end, spectra of a large sample of H II regions in LSB galaxies have been obtained. The H II regions provide powerful probes of the physical conditions in the galaxies in which they reside. The strong emission lines are more readily observable than the weak continuum, and line ratio diagnostics contain information about the chemical abundances in the gas and the stars which ionize it. These H II regions are almost always giant H II regions ionized by clusters of stars rather than individual stars.

The H II regions were identified in the disks of LSB galaxies from narrow band H$\alpha$ images (McGaugh 1992). The sample was selected from the lists of Schombert & Bothun (1988) and Schombert et al. (1992), and from the UGC (Nilson 1973). All of these galaxies have central surface brightnesses well below the canonical Freeman (1970) value of $\mu_0 = 21.65$ mag arcsec$^{-2}$, with the sample median being $\mu_0 = 23.4$ mag arcsec$^{-2}$ (McGaugh & Bothun 1993). A wide range of morphologies are present, and no single type can be considered representative (see images in McGaugh, Bothun & Schombert et al. 1993a). Nonetheless, it is not uncommon for the H II regions to be imbedded in a weak spiral pattern, though there are generally not enough H II regions along the arms to define them as in Sc galaxies.

It is important to realize that while these LSB galaxies are all disk systems of low surface brightness, they are *not* dwarf galaxies. Rather, they are normal size disk galaxies insofar as the scale length distribution of this sample is indistinguishable from that of the HSB spirals which define the Hubble sequence (McGaugh & Bothun 1993). This is an investigation of the properties of typical spiral size disks over a range of surface brightness, not of the relationship between star bursting dwarf galaxies and their presumably LSB progenitors (e.g., Tyson & Scalo 1988).

A *de facto* selection effect in a study of this kind is the presence of H II regions. While these are more common than might be expected in galaxies selected for low surface brightness, the sample can in no way be considered complete. It nevertheless provides the opportunity to study the physical properties of an interesting if not exhaustive portion of the LSB galaxy population.

## 2. OBSERVATIONS AND REDUCTIONS

Optical spectra of 71 H II regions in 22 LSB galaxies were obtained over the course of several observing seasons from September 1989 through February 1992. Blue spectra covering the 3500 – 5400 Å range were acquired with the Kitt Peak 2.1 m telescope and Gold Camera Spectrograph using a TI 800 x 800 CCD as the detector. Red spectra covering 5700 – 7000 Å were obtained with the Hiltner 2.4 m telescope, Mk III spectrograph, and TI-4849 CCD (Luppino 1989) of the MDM[1] Observatory. Long baseline spectra covering 4500 – 7300 Å were also obtained at MDM Observatory with this equipment.

Since LSB galaxies and their H II regions generally cannot be seen with the finder cameras of these telescopes, the slit was positioned by offsetting from nearby stars. The offsets were measured from the H$\alpha$ images used to identify the target H II regions. Slit angles were chosen to maximize the number of H II regions which could be observed simultaneously. To minimize the impact of differential atmospheric refraction and the uncertainty in the offset procedure, a wide (3.4") slit was used. This resulted in a resolution of $\sim 11$ Å for the blue spectra, $\sim 8$ Å for the red spectra (adequate for splitting [N II] from H$\alpha$ and the [S II] doublet), and $\sim 15$ Å for the long baseline spectra. These latter overlap the red and blue spectra and were used to place them on the same relative flux scale. The linking spectra also provide a consistency check. Except in a few cases where the offset procedure obviously failed, the line *ratios* agree to within the errors expected from counting statistics and the read noise of the CCD detectors. Because the spectra were not always obtained under photometric conditions, the *fluxes* of individual lines observed multiple times were not always the same. Fortunately, this study requires only line ratios, and is unaffected by uncertainties in the absolute zero point of the flux scale.

---

[1] MDM Observatory is operated by the University of Michigan, Dartmouth College, and the Massachusetts Institute of Technology.

Data reduction was accomplished primarily in IRAF[2]. This began with subtraction of the bias level and dark current. The frames were flattened in the spectral direction with lamp flats, and the slit illumination was made uniform with twilight sky flats. Emission line lamps provided spectral calibration and two dimensional justification for optimal sky subtraction. Standard star observations were used to flux calibrate the object frames. This latter step also corrected for the spectral shape of the instrumental response. Apertures containing individual H II regions were extracted and stored as one dimensional spectra. Line fluxes were measured from these by using the FIGARO package to fit gaussians to the nebular emission lines. The data were subsequently reanalyzed with the improved gaussian fitting routines in IRAF, recovering the same results within the errors.

The data are presented in Tables 1 and 2. Table 1 contains Balmer emission line data and information derived therefrom. Column 1 contains the galaxy name. Galaxies beginning with 'F' are from the lists of Schombert & Bothun (1988) and Schombert et al. (1992), and are designated by sky survey field number. Galaxies beginning with 'U' are from the UGC. Column 2 contains the H II region specification (McGaugh 1992). Column 3 contains the observed H$\beta$ flux in units of $10^{-18}$ ergs cm$^{-2}$ s$^{-1}$. Because conditions were not always photometric, these numbers may not always be within the formal errors listed in Column 4, which include the internal contributions to the uncertainty (the shot noise in the line and continuum and the read noise of the detector), but no estimate of the uncertainty in the flux calibration. Columns 5, 6, and 7 contain the observed equivalent width in Å of the Balmer lines H$\gamma$, H$\beta$, and H$\alpha$, respectively. Being a measure of the line relative to the surrounding continuum, these are not subject to uncertainties in the zero point caused by nonphotometric conditions. Indeed, the greatest uncertainty here is in the placement of the low continuum level. Column 8 contains the reddening coefficient $c$, related to the more familiar $E(B-V)$ by $E(B-V) = 0.78c$. The error in $c$ is listed in Column 9.

The reddening coefficient is determined from the interstellar extinction curve of Savage & Mathis (1979) with the observed Balmer decrements by assuming Case B ratios for a $T_e = 10,000\,K$ nebula in the low density limit. This is an adequate assumption, as the ratios of the [S II] lines generally indicate the low density limit, and the Balmer decrement is not very sensitive to either density or electron temperature. In a few cases of zero reddening and hot nebulae, modestly negative reddenings are derived. These are set to zero in the subsequent analysis.

---

[2] The Image Reduction and Analysis Facility (IRAF) is distributed by the Association of Universities for Research in Astronomy, Inc., under contract to the National Science Foundation.

The derived value of the reddening coefficient is sensitive to the presence of Balmer absorption in the underlying stellar continuum. This alters the Balmer decrement in the same sense as the reddening, so the latter will be overestimated if absorption is present but the departure from the Case B ratio is attributed entirely to interstellar reddening. The amount of absorption in any single object is very difficult to estimate, as the solution is always unconstrained. That is, observation of $n$ Balmer lines gives $n-1$ independent decrements to constrain $n+1$ variables (the $n$ absorption equivalent widths plus the reddening). This is commonly approached by assuming that the absorption equivalent widths are the same (McCall et al. 1985), or by constructing population synthesis models which are adjusted to subtract away the absorption (see e.g., Olofsson 1989). While the former approach may be an oversimplification, the latter requires specification of many potentially degenerate parameters (Silva 1991) such as the shape of the star formation history, the age of the star formation event, and the IMF. Such procedures may introduce more systematic uncertainties than they remove.

Though it is difficult to estimate the amount of absorption in individual spectra, the entire sample can be treated in a statistical sense by varying an assumed amount of absorption. In order to avoid the reddenings becoming negative, and introducing a spurious trend of reddening with emission equivalent width, the LSB data require $W_\lambda^{abs} < 4$ Å, and suggest $1 < W_\lambda^{abs} < 3$ Å. Though large amounts of absorption cannot be ruled out in a few individual cases (such as H II region A2 in F746–1 and S1A2 in UGC 6151), it appears to be $\sim 2$ Å in most LSB H II regions. This is the value found in the H II regions of HSB spirals (McCall et al. 1985; Oey & Kennicutt 1993), so in order to compute the reddening it is assumed that $W_\lambda^{abs} = 2$ Å. Variation from the assumed value is a fundamental systematic uncertainty in this work and all like it. However, the magnitude of the effect is small for the statistically allowed range of absorption, and a more serious (and basic) assumption may be that of the universal applicability of the standard extinction curve.

Table 2 contains the observed fluxes of the important nebular lines. Galaxies and H II regions are specified as per Table 1, followed by the linear intensities of the indicated lines relative to H$\beta$ (before correction for absorption or reddening) and their uncertainties. The uncertainty in the line ratios includes that in both the relevant line and in H$\beta$. In the case of object A2 in UGC 5709 and S3A1[3] in F568–6, H$\beta$ is only marginally detected so the listed error is large, but the ratios of the other lines to one another are rather more certain. For important marginal detections and significant nondetections, an upper limit denoted by a less than sign is given. It should also be noted that the measurements of [O III] $\lambda$4959 and $\lambda$5007 are not strictly independent, as they were initially constrained to be in their theoretical ratio. This constraint on the fit was subsequently relaxed, but the fit to these lines usually did not change much.

## 3. ANALYSIS

Figure 1 shows an example H II region spectrum with the important nebular lines identified. Note that this is not a typical spectrum; it is one of the better obtained and is chosen to illustrate the relevant lines. Many H II regions in this sample are too faint for any but the strongest lines to be measured with confidence. In particular, while the [O III] $\lambda$4363 line is sometimes visible, it is generally not observed with sufficiently high signal to noise for a reliable electron temperature to be determined. This makes the uncertainties in the abundances derived from the nebular spectra using standard methods (Osterbrock 1989) intolerably high.

To overcome this difficulty, the empirical abundance indicating line ratio $R_{23} \equiv$ ([O II] $\lambda$3727 + [O III] $\lambda\lambda$4959, 5007)/H$\beta$ first discussed by Pagel et al. (1979), as calibrated by McGaugh (1991), is employed. An important aspect of this calibration is the recognition of the influence of the ionization parameter at low abundances. When this is considered, the $R_{23}$ method rivals the precision of the standard method when an accurate, direct measure of the electron temperature is available, but has the advantage of relying only on the strongest nebular lines.

### 3.1. *Oxygen Abundances from the Strong Line Method*

Figure 2 shows the reddening corrected data for LSB H II regions plotted on the model grid of McGaugh (1991). The oxygen abundance and volume averaged nebular ionization parameter <U> (essentially the ratio of ionizing photon density to particle density) can be extracted from this plot. While $R_{23}$ alone is sufficient for determining the oxygen abundance on the upper branch (i.e., the solid lines of constant abundance in Figure 2 are nearly vertical), this is not the case on the lower branch. Here $R_{23}$ depends on $<U>$ as well as on abundance, so these must be determined simultaneously. This is accomplished by plotting the ionization sensitive line ratio $O_{32} \equiv ([O\ III]\ \lambda\lambda 4959, 5007)/([O\ II]\ \lambda 3727)$ as the ordinate of Figure 2. While both $O_{32}$ and $R_{23}$ depend on both $<U>$ and oxygen abundance on the lower branch, together they contain sufficient information to determine both, and have the further advantage of involving the same strong lines observable in the same wavelength region.

---

[3] This is the only object in the sample which is not obviously an H II region. It is not prominent in $U$ (see image in McGaugh et al. 1993a), suggesting that local photoionization is not responsible for the observed emission. The spectrum is consistent with very low excitation photoionization or shock heating, so it may be that this object is a jet originating in or photoionized by the AGN in this giant galaxy (Bothun et al. 1990).

The nebular parameters (O/H, $<U>$) are extracted from Figure 2 by the projection of the positions of the data in the observed ($R_{23}$, $O_{32}$) plane into the (O/H, $<U>$) plane represented by the grid lines. The error in the derived parameters due to observational uncertainty is taken from the corresponding projection of the error ellipse. The error bars in Figure 2 are computed by propagating the uncertainty in the individual line measurements and the reddening through to the plotted quantity. Objects with excessively large errors are excluded from both the plot and further analysis. The uncertainty in the final values of (O/H, $<U>$) are taken from the combination (in quadrature) of the observational uncertainty determined in this fashion and the theoretical uncertainty in the actual position of the grid.

This uncertainty in the calibration varies with abundance. On the upper branch (the solid-lined surface of Figure 2), $R_{23}$ is accurate to $\sim 0.1$ dex in log(O/H) (which is the separation between the grid lines). The uncertainty in $<U>$ on the upper branch is $\sim 0.15$ dex. The uncertainty in both grows larger above solar abundance as many complicating effects come into play: opacity effects become significant in important cooling lines, the models become sensitive to assumed abundance ratios, and variations in density and ionization cease to be homologous (see Oey & Kennicutt 1993). Also, $<U>$ is likely to be systematically affected at large abundances by the lack of metal edges in the model stellar atmospheres, which affects the computed [O III]/[O II] ratio. Fortunately, virtually all LSB galaxies are on the lower branch (the dashed surface of Figure 2), as will be shown. Here these uncertainties are unimportant. The lower branch calibration is accurate to $\sim 0.05$ dex in oxygen abundance and $\sim 0.1$ dex in ionization parameter. For those objects in which [O III] $\lambda 4363$ is observed, the two methods are in agreement within the errors, though the $R_{23}$ method tends to give slightly higher abundances. A complete discussion of the uncertainties in the calibration and a comparison to abundances determined with a direct measurement of the electron temperature is given by McGaugh (1991).

A systematic uncertainty which is not included in the estimate of the error in O/H and <U> is the amount of underlying Balmer absorption. As discussed above, this is assumed to be $W_\lambda^{abs} = 2$ Å. The assumed value affects the flux of H$\beta$, and through it, the derived value of the reddening. Increasing the amount of absorption increases the inferred strength of H$\beta$ relative to the forbidden lines, pushing the data to the left in Figure 2. There is also a small upward component to this vector as [O III]/[O II] will increase as the reddening estimate drops. This motion is of course largest for those objects with the smallest observed emission equivalent widths. Varying the assumed value of the absorption over the allowed (0 − 4 Å) range does not move most of the data outside the error estimates determined above.

Physically, the empirical method works as an abundance indicator because of the response of the R$_{23}$ lines to electron temperature (Pagel et al. 1979). As the abundance of coolants drops, the nebula becomes hotter and the optical oxygen lines increase in strength as they bear a growing fraction of the cooling burden. This cannot continue indefinitely, as there must be a point where the oxygen line strengths begin to decrease because oxygen becomes scarce (Edmunds & Pagel 1984). The maximum in R$_{23}$ occurs around log(O/H) $\approx -3.6$ (roughly 30% of solar), where the lines of constant oxygen abundance in Figure 2 become closely bunched. This bunching severely limits the accuracy of the R$_{23}$ method for objects near the fold, so the abundance can only be determined to be within $\sim 0.2$ dex of log(O/H) = −3.6. Many H II regions inhabit this region of the R$_{23}$ − O$_{32}$ diagram, which causes a somewhat artificial crowding of values near log(O/H) = −3.6. This point should be kept in mind when interpreting abundances determined in this fashion.

An important result follows directly from casual inspection of Figure 2. A number of points fall to the right of the fold, outside the range occupied by the model grid. This grid is for H II regions ionized by stellar clusters with an IMF truncated at $M_u = 60 M_\odot$. A grid for $M_u = 100 M_\odot$ would be displaced slightly to the right, bringing the ridge line of the fold into agreement (within the errors) with almost all the data. This is a strong indication that hot ($T_* > 50,000\,K$), high mass stars *are* present in many low surface brightness galaxies. This is consistent with the often large ionizing luminosities implied by the H$\beta$ fluxes in Table 1. If $M_u$ had a relatively low value ($\sim 30 M_\odot$), satisfying the ionizing luminosity requirement would require so many stars that the restriction on optical surface brightness would be violated. Though the precise value of $M_u$ is model dependent, the spectra indicate stellar temperatures comparable to those in starbursting compact galaxies (Campbell 1988) which have been cited as evidence for a metallicity dependent IMF (Terlevich 1985). This would imply an IMF biased *towards* high mass stars in LSB galaxies, although other lines of evidence suggest, if anything, an IMF *lacking* in massive stars (Romanishin, Strom, & Strom 1983; Schombert et al. 1990). Consideration of the effects of metallicity on stellar temperature shows that a fixed IMF adequatley explains the observed trends in the spectra (McGaugh 1991), so *no* variation in the IMF is inferred.

That there is a fold in the grid results in an ambiguity. While $R_{23}$ and $O_{32}$ specify O/H and $<U>$ with good precision on both branches, they do not specify which branch is appropriate. That is, the surfaces represented by solid and dashed lines in Figure 2 overlap, so that the values of (O/H, $<U>$) corresponding to any observed ($R_{23}$, $O_{32}$) are twofold degenerate. This means that two very different H II regions might have identical ($R_{23}$, $O_{32}$).

Fortunately, the easily observable line [N II] $\lambda6583$ can resolve the degeneracy. It is strong at high abundance, and quite weak at low abundance. Skillman (1989) suggested using it in the form of the empirical abundance indicator [O III]/[N II] of Alloin et al. (1979). This line ratio varies monotonically with abundance, but is also sensitive to $<U>$. A superior ratio for this purpose is [N II]/[O II]. This also varies monotonically with oxygen abundance, but is much less sensitive to $<U>$, and is observed to form a very narrow sequence over a large range of metallicity (McCall et al. 1985 — see Figure 3). Indeed, [N II]/[O II] would be superior to $R_{23}$ as an abundance indicator *if* the N/O ratio were known *a priori*. This is not the case, as N/O varies in a complicated fashion with O/H (e.g., Matteucci 1986; Vila-Costas & Edmunds 1993). Nevertheless, it is adequate to clearly distinguish between the upper and lower branches. In Figure 3, the division between upper and lower branches (the fold in Figure 2) occurs around log([N II]/[O II]) $\approx -1$. H II regions in HSB spirals mostly have [N II]/[O II] equal to or larger than this value (i.e., are on the upper branch; see the data of McCall et al. 1985). In contrast, the majority of LSB H II regions have [N II]/[O II] weaker than this, indicating that they are on the lower, dashed surface of Figure 2. That the narrow sequence defined by the HSB data disperses as log([N II]/[O II]) decreases past $-1$ is expected from the fact that the abscissa $R_{23}$ is sensitive to $<U>$ on the lower branch, and also results from variation in the N/O ratio.

It is an interesting coincidence that the transition between HSB and LSB samples occurs approximately at the turnover in the $R_{23}$ relation. However, this cannot be physically significant. The turnover is defined by local nebular physics, independent of any knowledge of the type of galaxy in which the H II region is imbedded. Indeed, that the samples just overlap suggests that there is a continuum of development histories in disk galaxies, and that surface brightness selection effects place limits on the range of our awareness of these.

## 3.2. *Results*

The results of the analysis are presented in Table 3. Galaxies and H II regions are specified as before. Entries are logarithmic, with abundances by number relative to hydrogen. The uncertainties are determined as described above, with observational and calibration uncertainties combined in quadrature. For a few objects in the turnover region with large errors, an abundance is adopted which is the midpoint of the extent of the error bars on either branch. These tend to be slightly less than the $\sim -3.6$ fold in Figure 2 as the lines of constant abundance are more closely bunched on the lower branch surface.

Because of bad weather, not all galaxies that were observed in the blue (where the $R_{23}$ lines are) were also observed in the red (where the [N II] line is), and in a few cases the observed limit on [N II] is not sufficiently restrictive to resolve the branch ambiguity. If no other indicator (such as the presence of [O III] $\lambda 4363$) of the appropriate branch is available, then the abundances determined from $(R_{23}, O_{32})$ remain ambiguous. Table 4 contains the possible oxygen abundances for those galaxies for which this is the case. Only one Balmer line is observed in most of these objects, so the reddening is assumed to be equal to the galactic value. A large (0.5 in $c$) random error is subsumed in the error estimate, but there is obviously no guarantee that the actual abundance determination is not more seriously affected.

The results presented in Tables 3 and 4 are plotted in Figure 4. Following the results for the galaxies for which the ambiguity is resolved, the objects in Table 4 are plotted (with different symbols) under the assumption that a lower branch assignment is appropriate. The inclusion of these objects has no bearing on the conclusions.

There is a widespread distribution in (O/H, <U>) in Figure 4 which is a direct consequence of that in the observed $(R_{23}, O_{32})$ plane. While there is clearly no tight correlation between O/H and <U> as implied by observations of H II regions in HSB galaxies (e.g., Dopita & Evans 1986), an *envelope* of maximum <U> which decreases with increasing O/H seems to be present (cf. Campbell 1988). Such an envelope would induce an apparent O/H–<U> correlation if only the highest surface brightness H II regions are selected for observation, as these will automatically be those with the highest <U> at a given metallicity. This suggests that the correlation observed in HSB galaxies is a selection effect, as argued by McGaugh (1991) on theoretical grounds.

That many values of <U> exist at any given O/H is expected from simple evolutionary considerations: as an H II region ages, the stars providing the ionizing luminosity will evolve and fade, and perhaps disperse the gas through the actions of stellar winds and supernovae. Both of these will reduce <U>, while O/H will not increase noticeably except at very low abundances (Kunth & Sargent 1986). Certainly, there is observational evidence that evolution is relevant (Campbell 1988, von Hippel & Bothun 1990, Caldwell et al. 1991). Hence one expects an unbiased sample to contain many <U>, not just those along the envelope. This predicts (McGaugh 1992) that studies to faint levels within HSB galaxies should find that H II regions fill in the area below the envelope; i.e., they would have a distribution similar to that in Figure 4. This is confirmed in the case of M101 by the data of Scowen, Dufour, & Hester (1992). However, evolution would appear not to be the only relevant effect, as Scowen et al. (1992) find examples of low surface brightness H II regions which are also large and luminous.

The significance of the envelope is unclear. If evolution is the primary cause of the distribution in in O/H–<U>, then it suggests that star clusters which become the ionizing sources for extragalactic H II regions form with the same typical mass and density that will result in the <U> value along the envelope for the appropriate metallicity. This would be an important clue to the metallicity effects on the star formation process. However, it is also possible that the clusters form over a wider variety of conditions rather than evolving to the various observed <U>. If this is the case, then the envelope would represent some limiting factor on the star formation process rather than being a universal result thereof. Certainly, the envelope represents some metallicity dependent threshold which warrants further examination.

The lack of high metallicity, high ionization H II regions that the envelope represents strongly suggests that the majority of objects with ambiguous abundance determinations are in fact on the lower branch. If not, these objects would lie in a region of the O/H–<U> plane not occupied by any other known extragalactic H II regions (Campbell 1988, Diaz et al. 1991), including those in LSB galaxies for which the branch assignment is certain. The only exceptions are F558–1 and UGC 6614, which have sufficiently low <U> that they do not exceed the envelope for either branch assignment, and so persist in being ambiguous (these are excluded from Figure 4). UGC 6614 is more similar in size and morphology to the giant Malin 2 (F568–6; Bothun et al. 1990) than to the rest of the objects in the sample. If this is an indicator, then it may belong on the upper branch with Malin 2. The choice of branch makes an enormous difference in UGC 6614: it is either the most metal poor or most metal rich object in the sample. The ambiguity in this important case needs to be resolved by further observation.

## 4. LOCAL CONDITIONS

The H II regions provide powerful probes of the local conditions where star formation is occurring. This is of considerable interest in the more pristine systems, which may be as close to the first epoch of star formation as is observable. The mode of star formation in LSB galaxies is in some way different from that in systems which result in high surface brightnesses. In particular, the strict limits placed on molecular emission from LSB galaxies (Schombert et al. 1990) and their low H I column densities (van der Hulst et al. 1993, McGaugh et al. 1993b) imply that the formation of molecular clouds may be impaired, and thus may not be the sites of star formation as is usually presumed to be the case. An important constituent of molecular stellar cocoons is dust, which must be present in substantial quantities to shield the molecules from the interstellar radiation field, and provide sites for the formation of molecular hydrogen. The reddening towards H II regions, usually high in HSB disks (Kennicutt 1983), provides an indication of the available dust content.

A histogram of the dust content as measured by the reddening $E(B-V)$ along the line of sight to the H II regions determined from the Balmer decrements and corrected for reddening due to our own galaxy (Burstein & Heiles 1984) is presented Figure 5. Also shown is the data of McCall et al. (1985) for Sc galaxies. Again, selection effects make the quantitative interpretation of the distribution in this histogram problematic. The two distributions do appear similar, with some a tendency for LSB galaxies to contain less dust. Nevertheless, there appears to be dust in at least some LSB galaxies which could potentially be associated with molecular material.

The lack of CO detections by Schombert et al. (1990) could result either from a real lack of molecular gas, or from the breakdown of the $CO/H_2$ conversion at low metallicity (Maloney & Black 1988; see also Adler, Allen, & Lo 1991). The limits placed by Schombert et al. (1990) are sufficiently strict that it is unlikely that the low metallicities of LSB galaxies are the only cause of the lack of detections. Indeed, Sage et al. (1992) find no correlation between $L(CO)/M(HI)$ and metallicity, in which case the large H I masses of LSB galaxies should lead to easily detectable molecular gas if the latter is present in "normal" relative quantities.

While the presence of massive amounts of $H_2$ cannot be ruled out due to the uncertainty in the conversion factor, it seems likely that LSB galaxies are either devoid of molecular gas, or possess modest amounts likely to be distributed in a cloud mass spectrum different from that in HSB galaxies as a result of the low H I column densities.

Any impairment of molecular cloud formation could in turn impair star formation, and lead to the current, relatively unevolved state of LSB galaxies. This supports the notion (van der Hulst et al. 1987, Kennicutt 1989, Bothun et al. 1990) that the surface density of H I is the fundamental physical quantity which controls the rate of disk evolution. Also, it should be noted that if LSB galaxies are indeed devoid of molecular gas, then the observed star formation is proceeding without the benefit of molecular clouds in sites fundamentally different from those observed in our own galaxy.

Not only can star formation proceed without apparent molecular clouds, it also seems able to proceed before dust is produced. Figure 6 shows the reddening towards H II regions in LSB galaxies as a function of abundance. A trend of dust content increasing with metallicity is apparent (see Campbell, Terlevich, & Melnick 1986 for a similar result in BCGs). That there is a positive trend indicates that dust, like the heavy elements of which it is composed, is a stellar product which exists only in trace amounts in the least evolved systems. Indeed, that there are so many systems with low reddening immediately suggests that the low surface brightness of these galaxies is *not* induced by large internal extinction. This is corroborated by the lack of IRAS detections of LSB galaxies (Schombert & Bothun 1988), which further suggests that they do not contain large amounts of gray dust.

## 5. GLOBAL EVOLUTION

A fundamental measure of the extent of evolution that a galaxy has undergone is the degree to which it has converted the primordial gas from which it formed into stars and, through them, to metals. The latter is traditionally quantified as the metallicity $Z$, or mass fraction in elements heavier than hydrogen and helium. However, only within the solar system is this quantity actually measured. In studies of stars, what is usually quoted is [Fe/H], the abundance of iron relative to the solar level. The conversion to $Z$ depends on the relative abundance of all other elements to iron, usually *assumed* to be solar. This turns out to be a very bad assumption for low metallicity stars in our own galaxy, where [Fe/O] is substantially subsolar (Abia & Rebolo 1989). This situation is complicated by the extreme variations that can occur in the iron to oxygen ratio depending on the star formation history of a galaxy (Gilmore & Wyse 1991). In other galaxies, the matter is further complicated in that studies of stellar metallicities are usually based on yet another element, usually magnesium (e.g., Thomsen & Baum 1987), while studies of the gas phase metallicity as probed by H II regions (such as this one) report the oxygen abundance. These differences, together with the fact that galactic absorption line strengths reflect the luminosity weighted average metallicity of a composite stellar population, while H II region abundances probe the local gas phase abundance, makes intercomparison of the results from these various methods problematic.

Of the possible approaches, the data presented here are probably best suited to the purpose at hand. The current gas phase abundance should represent the highest metallicity reached by contributions from all generations of stars. Oxygen is the most abundant of the metals, and provides the plurality of mass therein. As such, it is the best tracer of the metallicity $Z$. Also, it is produced in quantity exclusively by massive ($M \gtrsim 10 M_\odot$) stars. Thus it originates in a well defined population and is returned to the interstellar medium on short timescales. In contrast, iron is contributed by several sources on a variety of timescales (Wheeler, Sneden, & Truran 1989).

Figure 7 shows the histogram of abundances for LSB disk galaxies. The prominent peak at $\log(O/H) = -3.6$ results in part from the uncertainty in the $R_{23}$ method at this abundance. There really are a fair number of H II regions *centered* around $\log(O/H) = -3.6$, but the larger errors at this abundance contribute to the appearance of an artificially sharp spike. This and selection effects make the quantitative interpretation of Figure 7 difficult.

Nevertheless, several points can be made. First, LSB systems cover a wide range in metallicity, over an order of magnitude in O/H. No value is obviously preferred, as it is unclear if the distribution is truly bimodal, or merely broad with the spike inducing an artificial appearance of bimodality. A broad spectrum of slow enrichment timescales appears to be represented. This is not surprising given the diversity of size and morphology represented in this sample, and suggests that many evolutionary histories are possible which do not lead to the high surface brightness spiral galaxies which are considered "normal."

Second, the LSB galaxies with the lowest metallicities are as low as any known extragalactic objects with the exception of I Zw 18 and SBS 0335 − 052 (Izotov et al. 1990). Kunth & Sargent (1986) showed that an abundance of $\log(O/H) \approx -4.3$ can be reached in a very short time (a few million years) as the result of production in the first stars to explode as supernovae. They suggest that this may be the reason that extragalactic H II regions more metal poor than this are extremely rare: as soon as star formation commences, even primordial gas will be enriched to this level very rapidly. Thus it is possible that objects with $\log(O/H) \lesssim -4.3$ are undergoing their first episode of star formation. Indeed, based on abundance and color, Salzer et al. (1991) conclude that the stars producing the bulk of light of the optical component of the H I cloud in Virgo are extremely young. This object is a typical low surface brightness galaxy by the standards of this sample. Regardless of the age of the stellar population, it is clear that these galaxies are underevolved relative to their HSB counterparts, and that selecting for low surface brightness is as effective as other means of discovering low metallicity objects (cf. Skillman et al. 1989).

That these galaxies are relatively unevolved is the third, and most important, point. Virtually all LSB galaxies have substantially subsolar abundances, despite being comparable to HSB spirals in size and mass. The only galaxies containing H II regions with $\log(\mathrm{O/H}) > -3.6$ are UGC 5709 and F568–6 (Malin 2). UGC 5709 is relatively high surface brightness ($\mu_0 = 22.49$ mag arcsec$^{-2}$), brighter than the sample median by nearly a magnitude (McGaugh 1992). Considering its extremely large size, Malin 2 may be fundamentally different from the other galaxies in this sample. Excluding these two galaxies leaves only those with $\log(\mathrm{O/H}) < -3.6$, or $Z < 0.3\ Z_\odot$

## 6. PHYSICAL RELATIONS

Since the metallicity is a measure of the degree of evolution of a galaxy, it is of considerable interest to see if there are any correlations between the measured abundances and other physical properties. Previous work on (predominantly HSB) disk galaxies have indicated correlations between metallicity and mass (e.g., Garnett & Shields 1987, Vila-Costas & Edmunds 1992, Oey & Kennicutt 1993), luminosity (Skillman et al. 1989 and references therein), and surface brightness (Webster & Smith 1983). Relations like these for the present sample of LSB galaxies are examined here, with relevant data taken from McGaugh (1992).

An important caveat here is that the H II region measurements provide inherently local measures of the metallicity, and there are real abundance variations across the disks of galaxies. It may therefore be inappropriate to compare locally determined abundances with global properties. To address this difficulty, varied attempts have been made to define a globally characteristic metallicity, such as the abundance at some standard radius (Garnett & Shields 1987) or a fit to the abundance gradient observed across the disk (Vila-Costas & Edmunds 1992). Unfortunately, there generally are too few H II regions in LSB galaxies to readily apply these methods. There are indications that there is little variation in some objects, and perhaps steep gradients in others. This will be investigated in greater detail in the future, together with how abundance varies with locally defined quantities (e.g., Phillipps & Edmunds 1991). For now it seems best simply to present the available data.

## 6.1. *Size*

Figure 8 shows the oxygen abundances plotted against the exponential scale lengths of the disks. These LSB galaxies cover a range of over an order of magnitude in size, most being comparable to the HSB spirals which define the Hubble sequence (McGaugh & Bothun 1993). However, they have very much lower abundances, with only the two galaxies UGC 5709 and Malin 2 discussed above having H II regions in the area log(O/H) $> -3.6$ occupied by HSB spirals. Thus disk galaxies can be found which exist over a range of metallicities, and hence degrees of development, at *all* sizes.

## 6.2. *Surface Brightness*

Since selecting a sample for low surface brightness has resulted in the discovery of low metallicity systems, one might expect a correlation of metallicity with surface brightness. However, there is no obvious, tight correlation between the individual H II region abundances and the *central* surface brightnesses of the disks (Figure 9). There must be some dependence, or the selection of the sample would not have resulted in only low abundance objects. There may be some mean trend which the current data are too sparse to elucidate, or this may simply indicate the difficulty in comparing local and global quantities mentioned above. Nonetheless, it may be that the evolution of a localized portion of a stellar system is fundamentally a local phenomenon which proceeds without knowledge of the type of galaxy in which it is imbedded (cf. van der Hulst et al. 1987).

## 6.3. *Environment and Morphology*

As LSB galaxies are predominantly (though not exclusively) late type galaxies (see Schombert et al. 1992; McGaugh et al. 1993a), then to the extent that this is meaningful it may be fair to say that metallicity and morphology are correlated. However, this is only true to the same degree as it is for central surface brightness — the two go together, and result in the selection of low metallicity objects. This does not, of course, guarantee a tight one to one correlation, and may only indicate a trend with substantial scatter.

In this same general sense, surface brightness is dependent on environment (Bothun et al. 1993; Mo, McGaugh & Bothun 1993) in the sense that LSB galaxies avoid regions of high galaxy density. Hence, more isolated galaxies are, on average, more metal poor. This result is complementary to that of Shields, Skillman, & Kennicutt (1991), who find an enhancement of metallicity in cluster spirals. However, Henry et al. (1992) dispute this, and argue that the obvious effects of cluster membership (e.g., gas stripping) do not have a serious impact on the chemical evolution of disks. Such processes are obviously unimportant in isolated LSB galaxies, so the trend of abundance with environment may point to some other underlying property, such as formation epoch (Mo et al. 1993, McGaugh et al. 1993b).

### 6.4. *Luminosity*

The relation between metallicity and luminosity is investigated in Figure 10. The oxygen abundances are plotted against the absolute magnitudes of the disk component *only*. A short ($H_0 = 100$ km s$^{-1}$Mpc$^{-1}$) distance scale is assumed. Lengthening the distance scale, or including the contribution of any bulge component (usually minor) would have the effect of sliding the data to the right in the diagram (i.e., brighter magnitudes at fixed oxygen abundance).

There is a great deal of scatter in this diagram. Indeed, if the brightest object (Malin 2) is excluded, there is no indication of any trend of abundance with luminosity. These LSB galaxies do not follow the $M_B$–$Z$ relation for dwarf irregulars found by Skillman et al. (1989), nor do they form a separate sequence of their own. This cannot be ascribed to the use of locally measured abundances, as the same method was used to define the relation. Galaxies exist over a wide range of properties, and metallicity–luminosity relations should be regarded with due consideration for the sample selection effects involved (cf. Bertola, Burstein, & Buson 1992).

### 6.5. *Mass*

The situation is similar for mass (Figure 11). No mass–metallicity relation exists for LSB disk galaxies, and there is little indication that metallicity increases at all with mass. There are objects comparable in mass to the Milky Way with substantially subsolar abundances. It is not obvious that LSB galaxies would follow the trends seen by Vila-Costas & Edmunds (1992) even if a similar analysis could be performed. Certainly, the giant Malin 2 is metal poor for its mass unless it has a very steep abundance gradient *and* there is a flattening of the trend defined by HSB spirals above $10^{11} M_\odot$.

Considering the selection effects involved in choosing both galaxies and individual H II regions, there remains much to be done to elucidate the relationship (if any) between metallicity and the various global properties of galaxies. Low surface brightness galaxies hold both a promise and a warning for such endeavors. Including them provides a long baseline over which the predominant effects may be distinguished. Sampling them at the low rate caused by selection effects skews our perception of the physical properties of galaxies.

# 7. CONCLUSIONS

Oxygen abundances in a sample of low surface brightness disk galaxies have been determined, yielding important clues to the evolutionary histories of these enigmatic objects. LSB galaxy H II regions are observed to contain massive ($M > 60 M_\odot$) stars, so no variation of the IMF with surface brightness is inferred, though the exact shape of the IMF is not constrained. The correlation between the nebular ionization parameter and metallicity observed in high surface brightness galaxies is not seen in this sample, but does seem to be present as an envelope of limiting maximum ionization at a given metallicity. This confirms the prediction by McGaugh (1991) that the apparent $<U>$–Z correlation arises from surface brightness selection effects.

LSB galaxies are low metallicity, slowly evolving systems. The low degree of chemical enrichment confirms the inference of Bothun et al. (1990) that LSB disks evolve slowly, and can remain quiescent for a Hubble time. This may result from their low H I column densities and lack of molecular gas. Indeed, the low H I column densities are likely to inhibit molecular cloud and star formation, and may be the physical attribute which determines the rate of galaxy evolution.

There are trends of metallicity with surface brightness and environment in the sense that isolated, LSB objects tend to be metal poor. This holds regardless of size, with some quite large galaxies having substantially subsolar oxygen abundances. No tight correlations are apparent between metallicity and global properties like luminosity and mass. Surface brightness selection effects can mask the true range of galaxy properties, and may cause the appearance of spurious correlations between physical properties. Recovery of the entire spectrum of galaxies and galaxy properties is essential to a comprehensive understanding of galaxy formation and evolution.

I am greatly indebted to Greg Bothun for his unflagging support for this project. It is a pleasure to acknowledge the insightful commentary and helpful input provided by him, Robert Kennicutt, Marshall McCall, and Jim Schombert. I am also grateful to the staffs of MDM and KPN Observatories for facilitating successful observing runs.

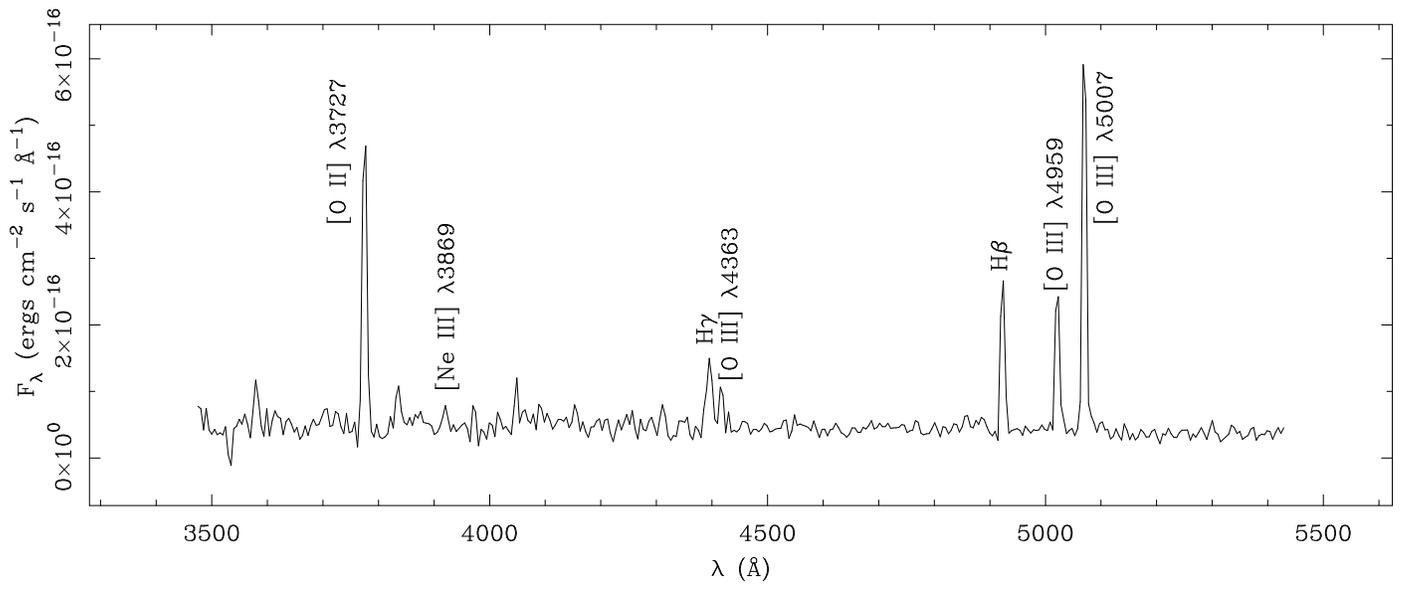
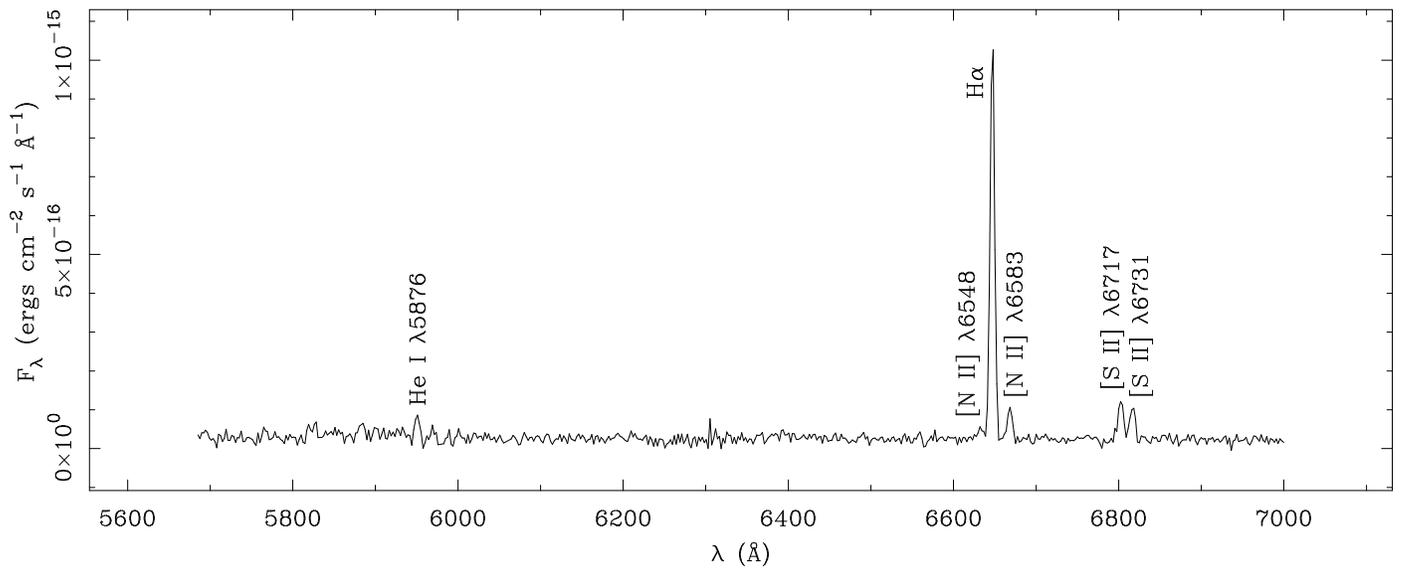

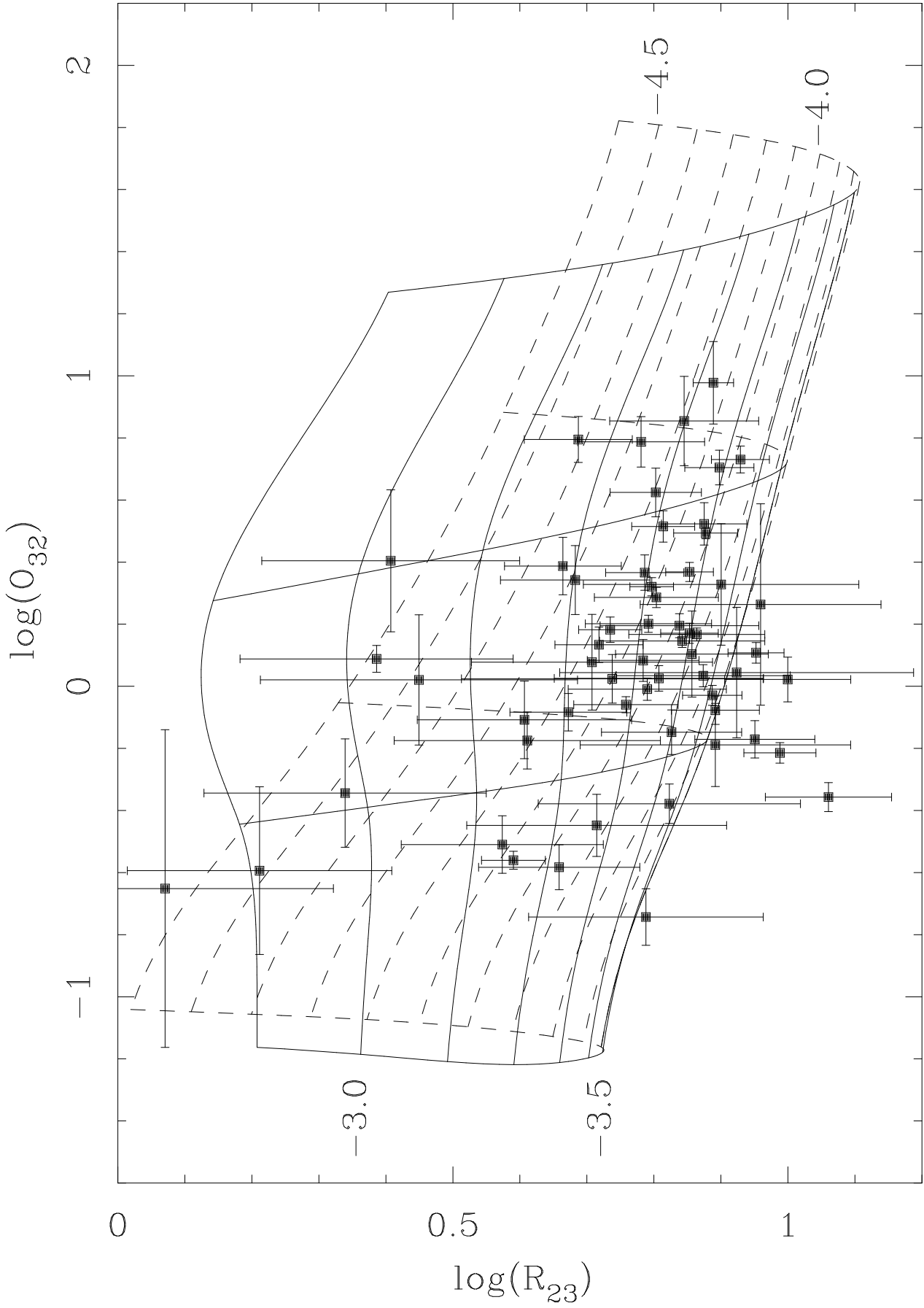

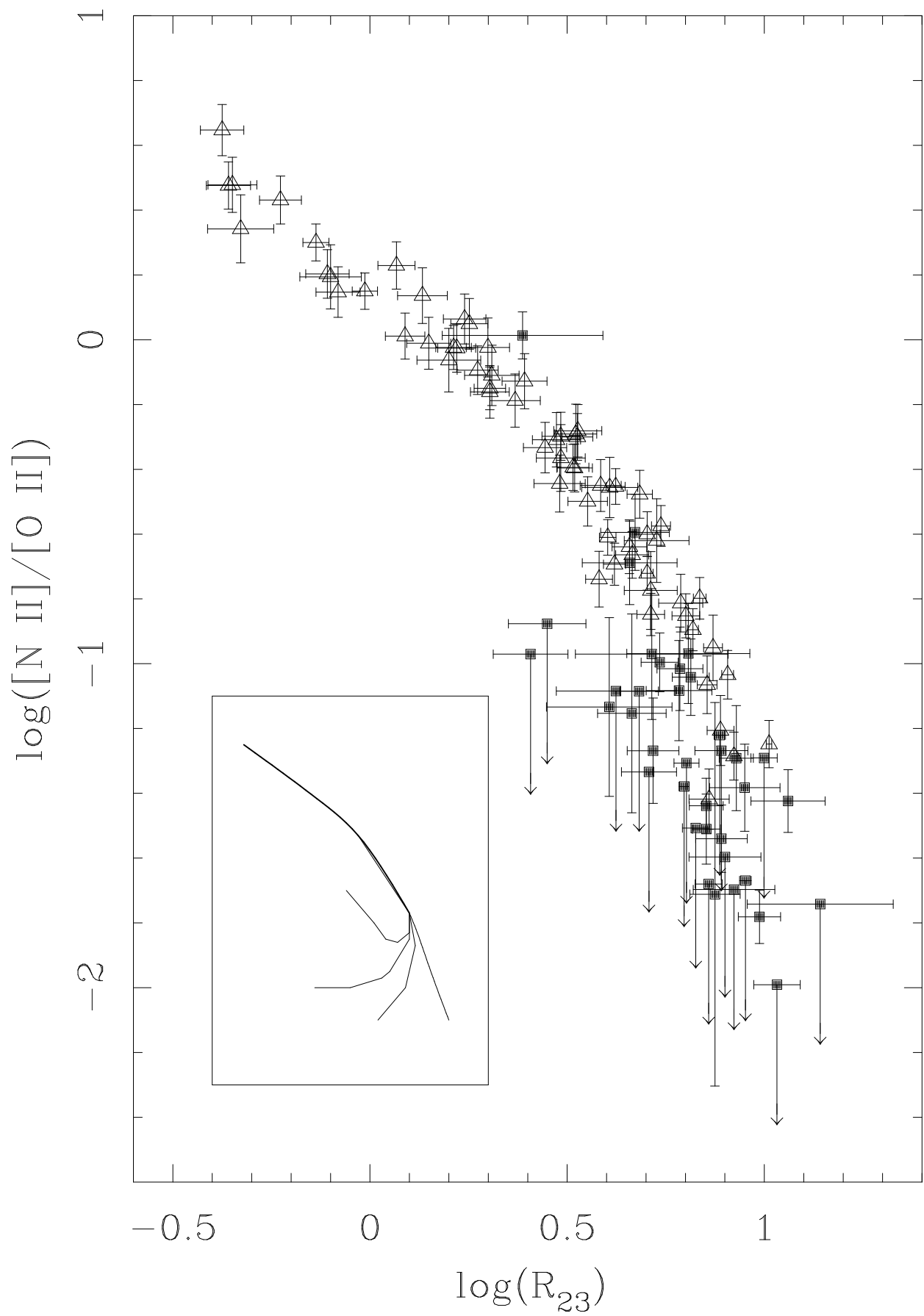

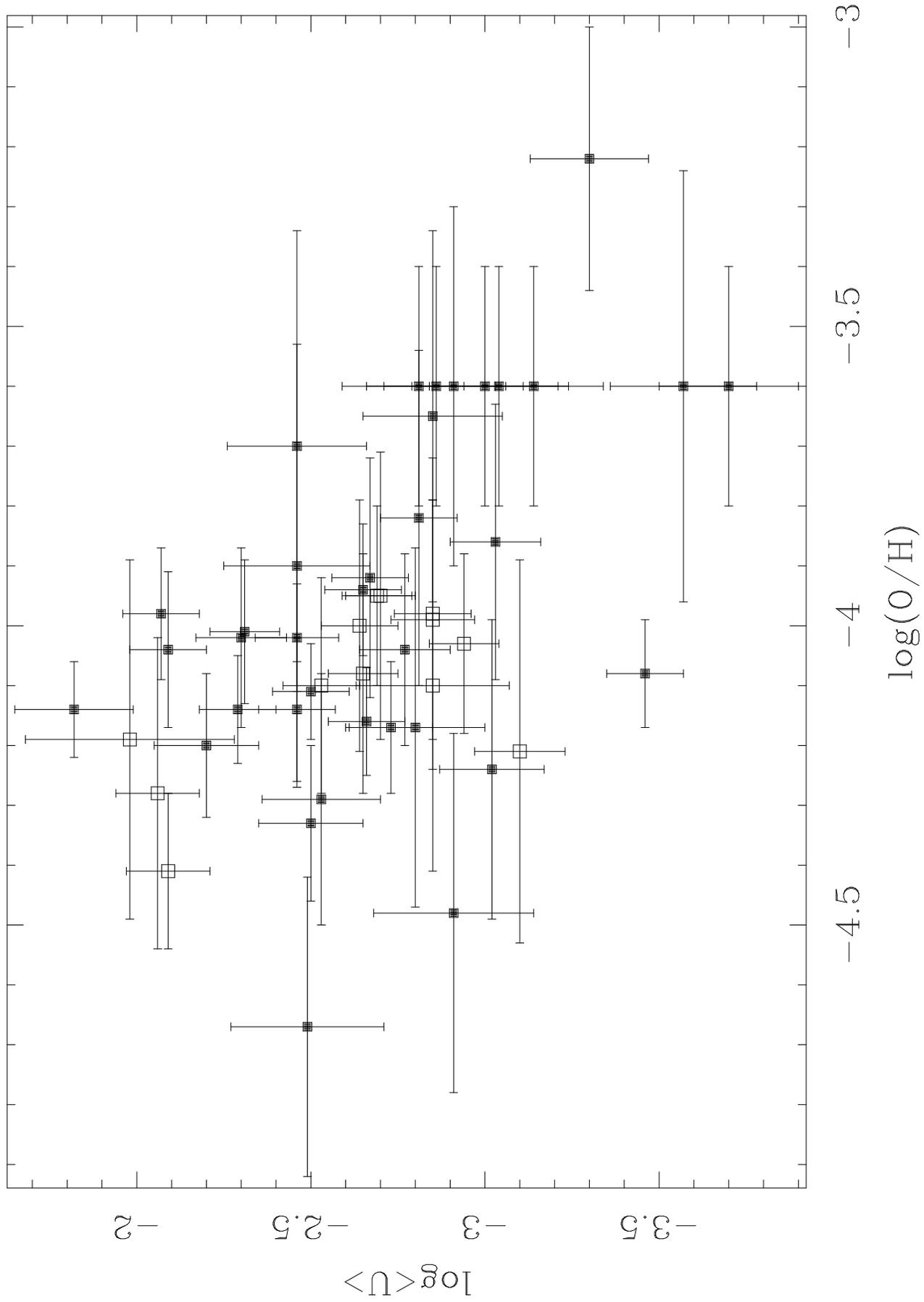

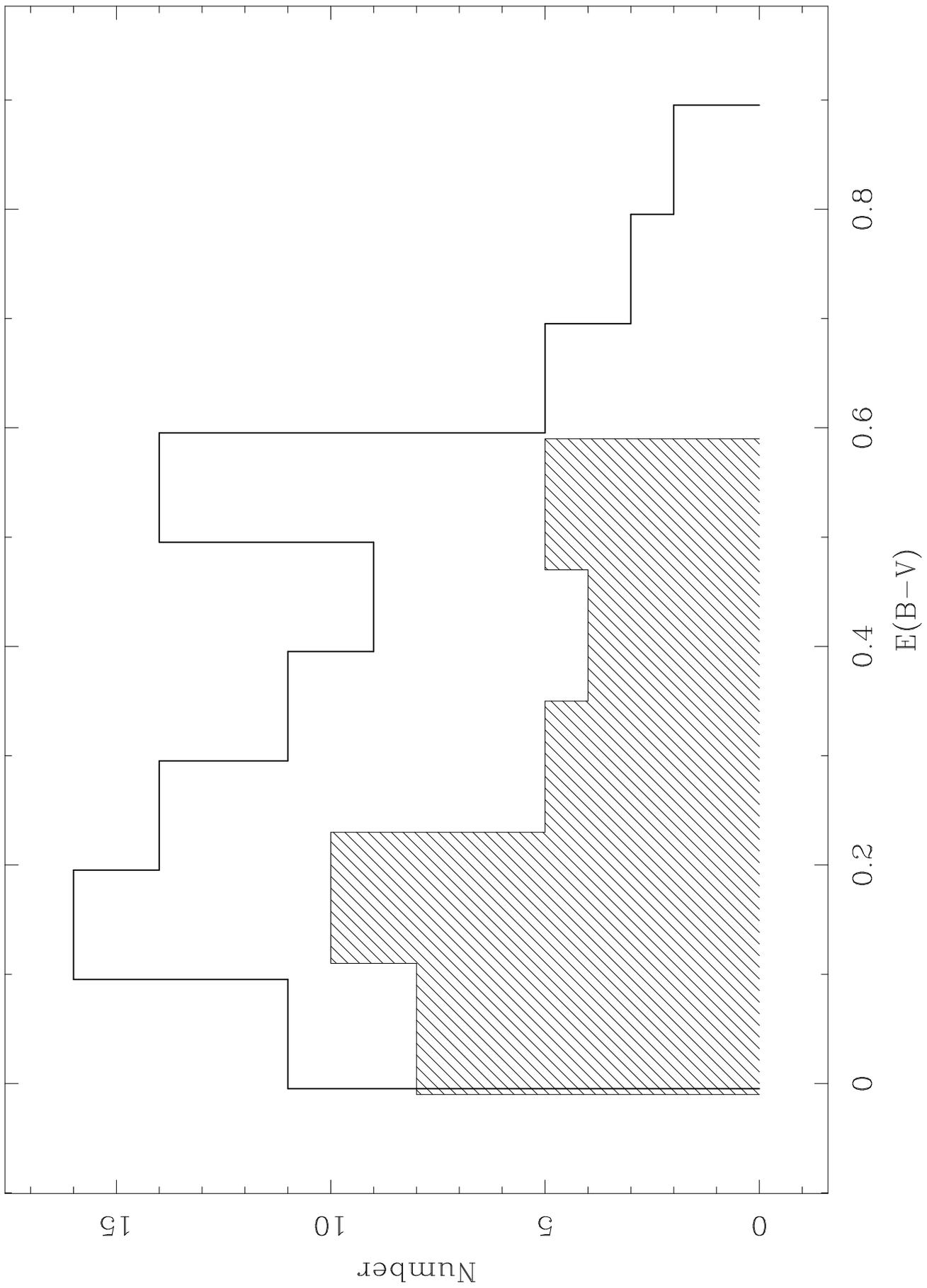

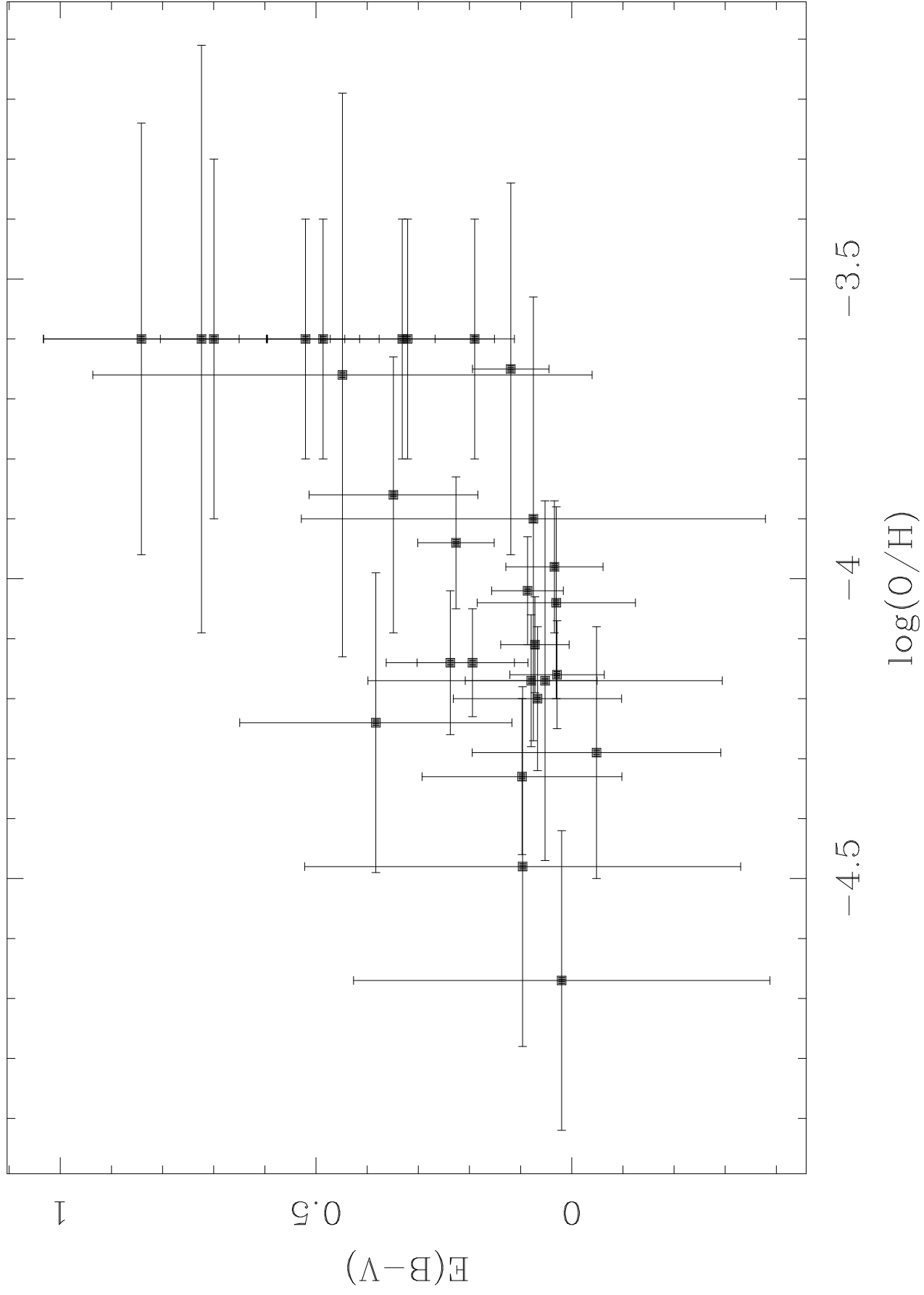

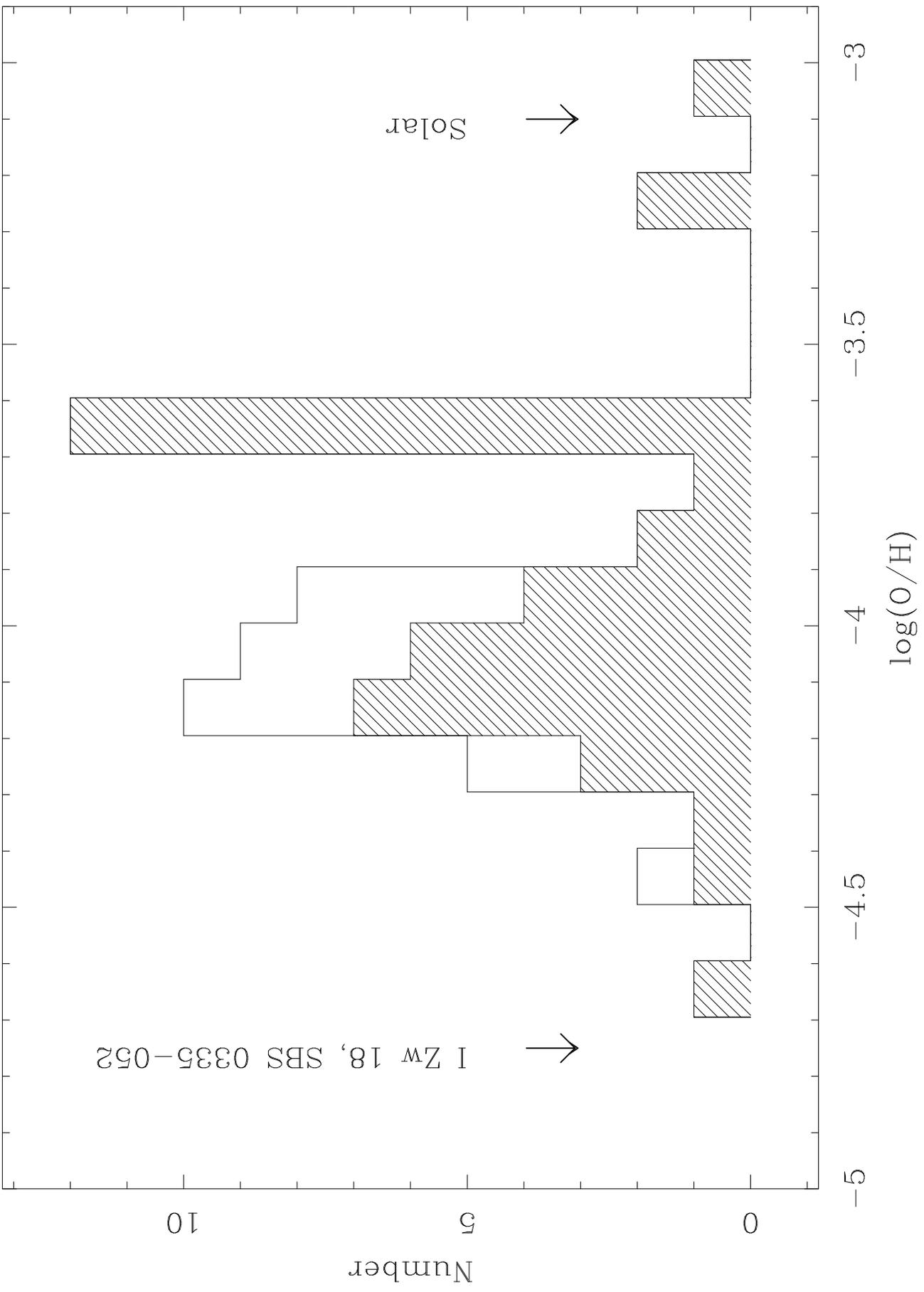

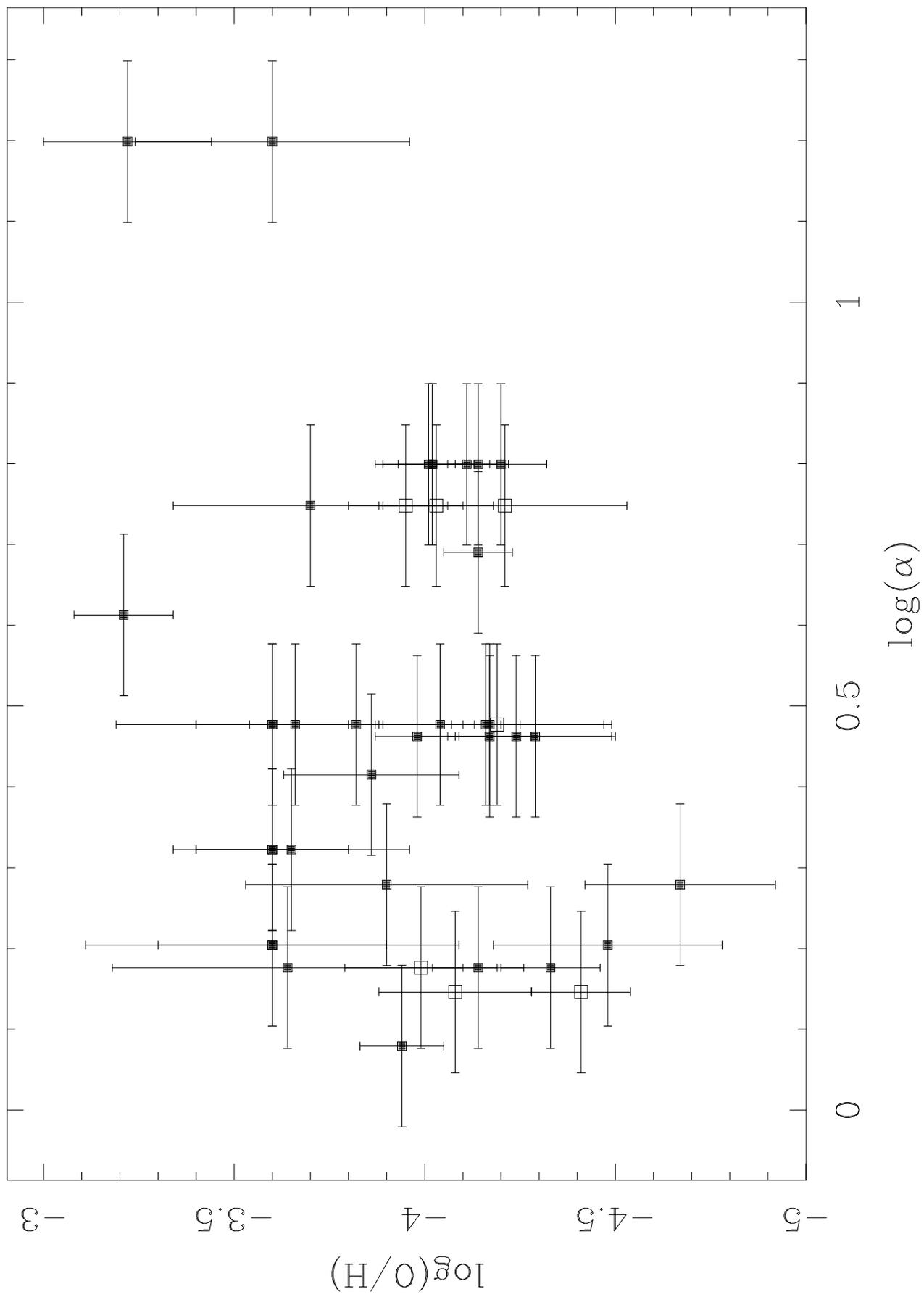

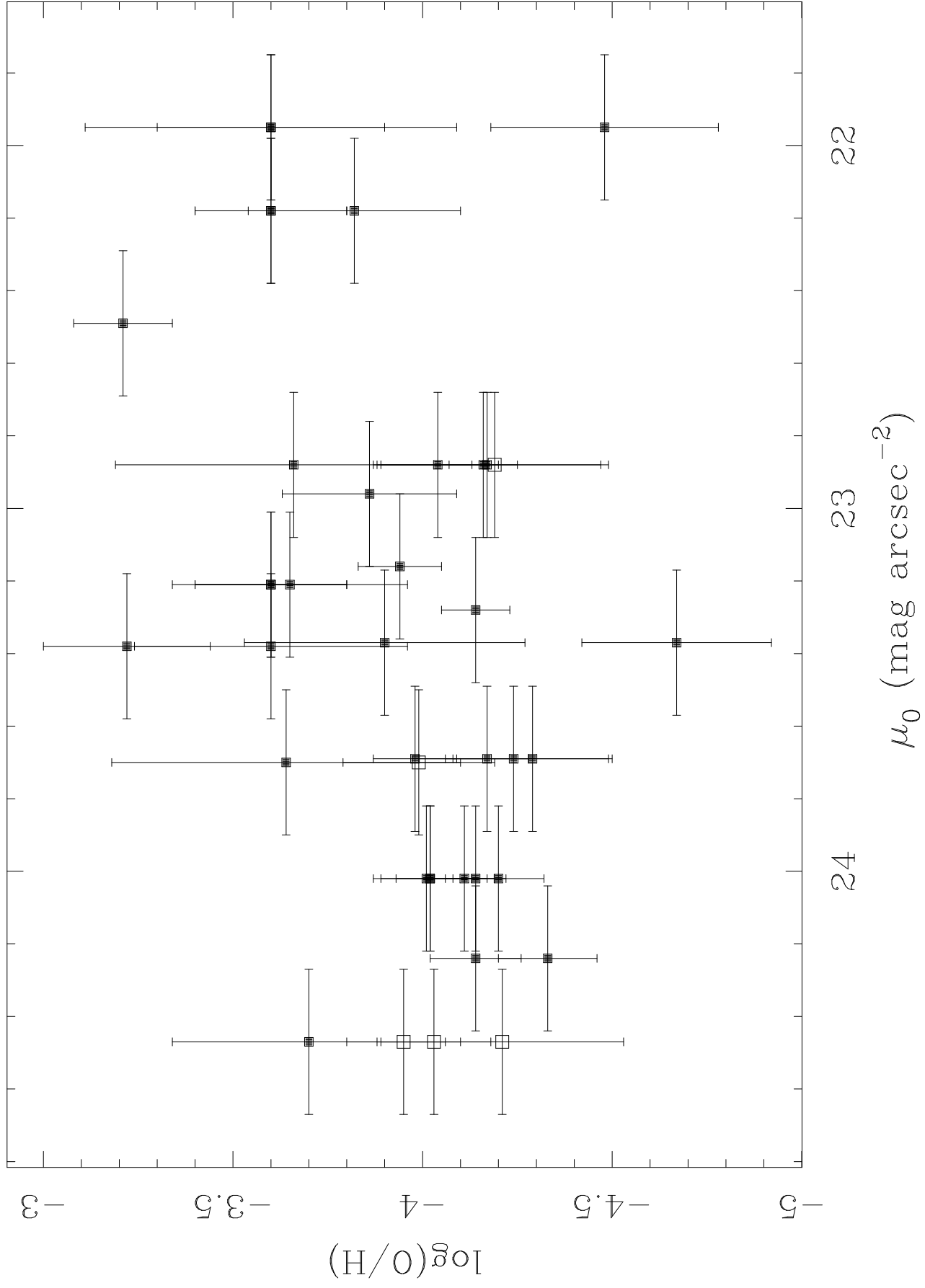

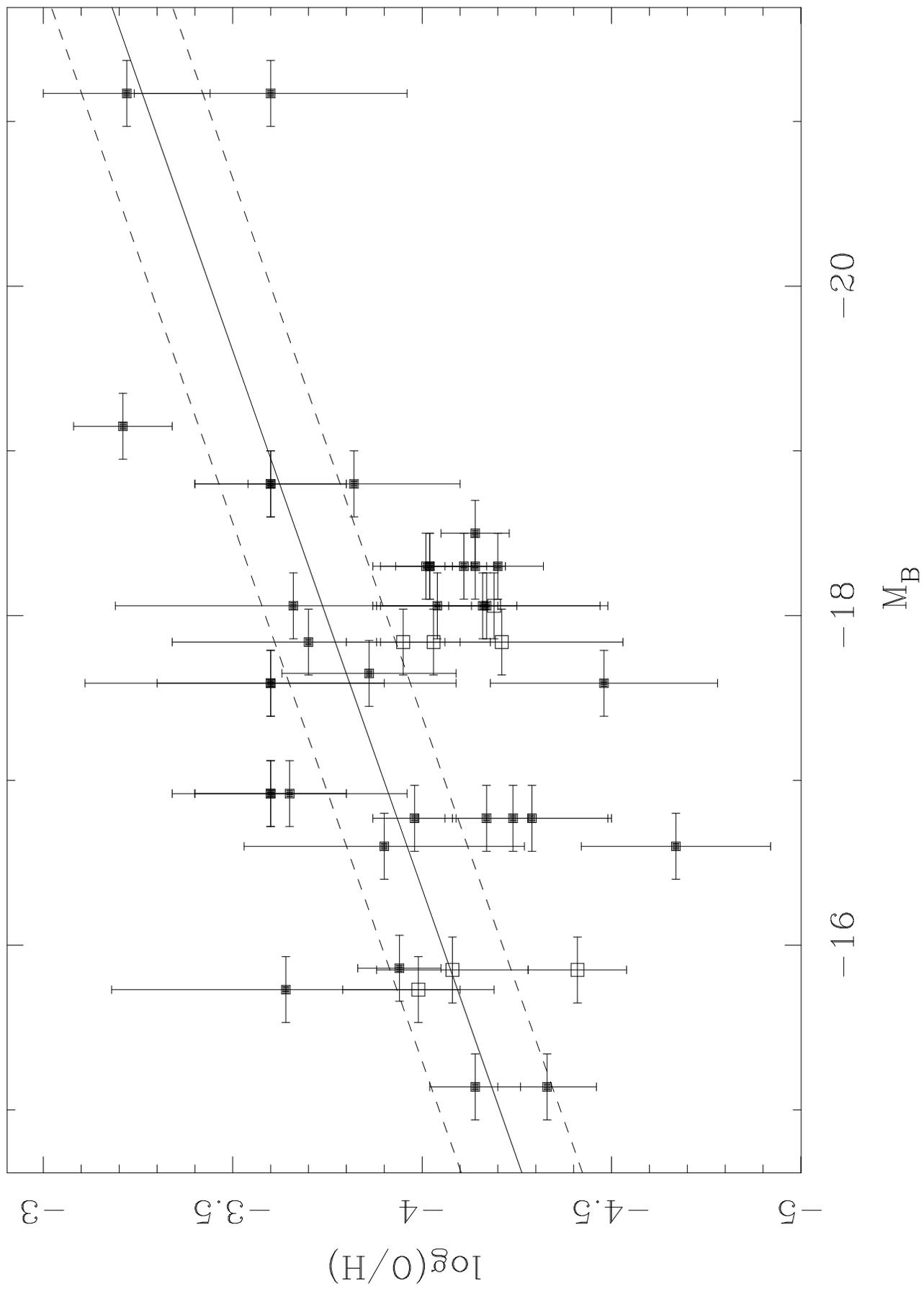

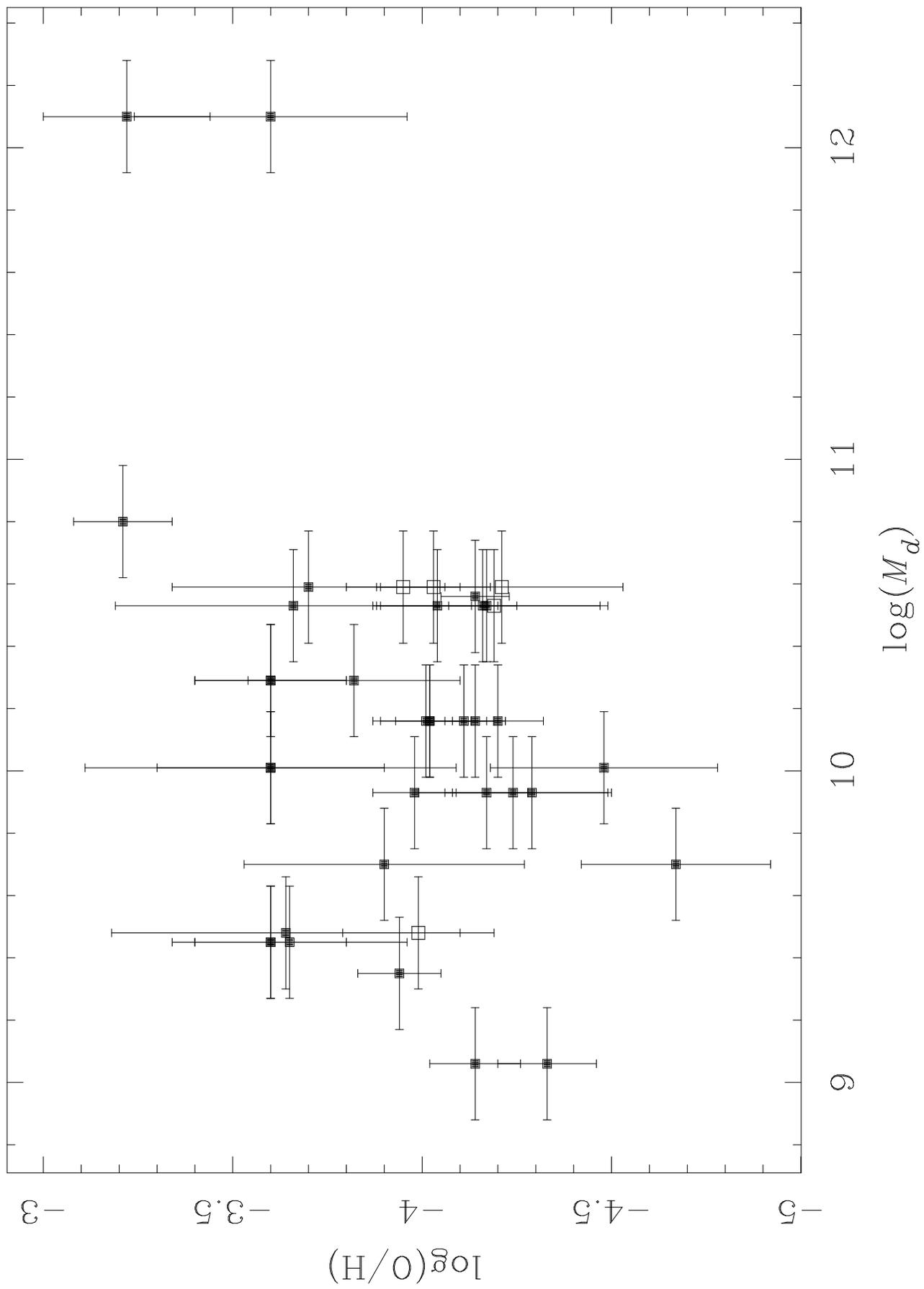

TABLE 1
BALMER LINE DATA

| Galaxy | H II | $F(H\beta)$ $10^{-18} erg\,cm^{-2}\,s^{-1}$ | $\sigma_{H\beta}$ | $W_\lambda(H\gamma)$ Å | $W_\lambda(H\beta)$ Å | $W_\lambda(H\alpha)$ Å | $c$ | $\sigma_c$ |
|---|---|---|---|---|---|---|---|---|
| F415-3 | A1 | 2964 | 94 | 16 | 35 | 113 | 0.383 | 0.096 |
| F469-2 | A1 | 1368 | 74 | 34 | 50 | 350 | 0.168 | 0.165 |
| F469-2 | A2 | 594 | 62 | 12 | 24 | 152 | 0.005 | 0.311 |
| F469-2 | A3 | 2123 | 84 | 21 | 39 | 259 | 0.110 | 0.121 |
| F469-2 | A4 | 515 | 60 | 18 | 32 | 152 | 0.558 | 0.341 |
| F530-3 | A1 | 1727 | 79 | 33 | 97 | 466 | 0.396 | 0.139 |
| F530-3 | A2 | 509 | 60 | 5 | 26 | 134 | 0.075 | 0.353 |
| F558-1 | A1 | 575 | 61 | … | 5 | 17 | 0.076 | 0.288 |
| F558-1 | A2 | 323 | 56 | … | 5 | … | … | … |
| F561-1 | A1 | 164 | 53 | … | 5 | 28 | −0.007 | 0.869 |
| F561-1 | A2 | 52 | 51 | … | … | 13 | … | … |
| F561-1 | A3 | 925 | 67 | … | 22 | 49 | 0.479 | 0.211 |
| F563-V1 | A1 | 276 | 56 | … | 13 | 49 | 0.112 | 0.581 |
| F563-V1 | A2 | 306 | 56 | … | 8 | 44 | 0.041 | 0.521 |
| F563-V2 | A1 | 294 | 56 | 5 | 12 | 32 | 0.129 | 0.546 |
| F563-V2 | A2 | 399 | 58 | 2 | 10 | 48 | 0.934 | 0.396 |
| F563-V2 | A3 | 1693 | 78 | 9 | 32 | 131 | 0.903 | 0.134 |
| F568-1 | A1 | 219 | 54 | … | … | … | … | … |
| F568-1 | A2 | 225 | 55 | … | 14 | … | … | … |
| F568-6 | S1A1 | 683 | 63 | … | 8 | 97 | 1.085 | 0.245 |
| F568-6 | S1A2 | 744 | 64 | … | 11 | 109 | 0.787 | 0.238 |
| F568-6 | S3A1 | 62 | 51 | … | … | 12 | … | … |
| F568-6 | S3A2 | 1030 | 69 | 5 | 19 | 171 | 0.535 | 0.192 |
| F568-6 | S3A3 | 140 | 53 | … | 14 | 118 | 0.908 | 1.011 |
| F577-V1 | A1 | 213 | 54 | … | 25 | 155 | 0.588 | 0.716 |
| F577-V1 | A2 | 128 | 53 | … | 7 | 41 | 1.377 | 1.085 |
| F577-V1 | A3 | 1157 | 71 | 11 | 28 | 164 | 0.428 | 0.180 |
| F577-V1 | A5 | 908 | 67 | 8 | 31 | 130 | 0.414 | 0.217 |
| F583-1 | A2 | 166 | 53 | … | 10 | 39 | 0.220 | 0.898 |
| F583-1 | A3 | 1502 | 76 | 14 | 47 | 125 | −0.004 | 0.157 |
| F583-5 | A1 | 753 | 64 | 5 | 13 | … | … | … |
| F583-5 | A2 | 802 | 65 | … | 43 | … | … | … |
| F585-3 | S1A1 | 880 | 66 | … | 9 | 152 | … | … |
| F585-3 | S1A2 | 1413 | 74 | … | 8 | 48 | … | … |
| F585-3 | S1A3 | 368 | 57 | … | 8 | 76 | … | … |
| F585-3 | S2A1 | 94 | 52 | … | 19 | … | … | … |
| F585-3 | S2A2 | 487 | 60 | … | 30 | … | … | … |
| F585-3 | S2A3 | 817 | 65 | … | 6 | … | … | … |
| F585-3 | S2A4 | 692 | 63 | … | 21 | … | … | … |
| F611-1 | A1 | 1391 | 74 | 20 | 46 | 303 | 0.339 | 0.160 |
| F611-1 | A2 | 788 | 65 | … | 53 | 982 | 0.159 | 0.250 |
| F746-1 | A1 | 2864 | 93 | 18 | 53 | 120 | 0.294 | 0.099 |
| F746-1 | A2 | 1524 | 76 | 3 | 12 | 47 | 0.674 | 0.140 |
| F746-1 | A3 | 433 | 59 | … | 25 | … | … | … |
| U 1230 | A1 | 392 | 58 | 5 | 45 | 264 | 0.195 | 0.443 |
| U 1230 | A2 | 1045 | 69 | 14 | 28 | 174 | 0.167 | 0.198 |
| U 1230 | A3 | 2218 | 85 | 21 | 51 | 249 | 0.165 | 0.118 |
| U 1230 | A4 | 242 | 55 | 8 | 11 | 58 | 0.702 | 0.625 |
| U 1230 | A5 | 527 | 60 | 17 | 33 | 101 | 0.131 | 0.344 |
| U 5675 | A1 | 263 | 55 | … | 9 | … | … | … |
| U 5675 | A3 | 466 | 59 | … | 17 | … | … | … |
| U 5709 | A1 | 493 | 60 | … | 2 | 34 | … | … |
| U 5709 | A2 | 99 | 52 | … | … | 57 | … | … |
| U 5709 | A3 | 1212 | 72 | … | 19 | 47 | 0.296 | 0.173 |
| U 6151 | S1A1 | 3073 | 96 | 16 | 47 | 148 | 0.152 | 0.095 |
| U 6151 | S1A2 | 2742 | 92 | 13 | 33 | 182 | 0.667 | 0.098 |
| U 6151 | S3A1 | 446 | 59 | … | 10 | … | −0.003 | 0.197 |
| U 6151 | S3A2 | 1839 | 80 | … | 56 | 424 | −0.004 | 0.055 |
| U 6614 | A1 | 298 | 56 | … | 7 | … | … | … |
| U 6614 | A2 | 258 | 55 | … | 9 | … | … | … |
| U 6614 | A3 | 244 | 55 | … | 9 | … | … | … |
| U 9024 | S1A1 | 1238 | 72 | 23 | 61 | … | 0.477 | 1.013 |
| U 9024 | S1A2 | 857 | 66 | … | 29 | … | … | … |
| U 9024 | S1A3 | 266 | 55 | … | 20 | … | … | … |
| U 9024 | S2A1 | 1369 | 74 | … | 54 | … | … | … |
| U12695 | S1A1 | 999 | 68 | 27 | 93 | 346 | 0.133 | 0.210 |
| U12695 | S1A2 | 3411 | 99 | 11 | 44 | 265 | 0.158 | 0.089 |
| U12695 | S1A3 | 1118 | 70 | 25 | 74 | 338 | −0.095 | 0.197 |
| U12695 | S2A1 | 3611 | 102 | 38 | 91 | … | 0.047 | 0.411 |
| U12695 | S2A2 | 3730 | 103 | 19 | 61 | 173 | 0.140 | 0.085 |
| U12695 | S2A3 | 1667 | 78 | 16 | 51 | … | 0.138 | 0.431 |

TABLE 2
Line Ratios

| Galaxy | H II | [O II] λ3727 | $\sigma_\lambda$ | [Ne III] λ3869 | $\sigma_\lambda$ | Hγ λ4341 | $\sigma_\lambda$ | [O III] λ4363 | $\sigma_\lambda$ | [O III] λ4959 | $\sigma_\lambda$ |
|---|---|---|---|---|---|---|---|---|---|---|---|
| F415-3 | A1 | 2.37 | 0.09 | 0.45 | 0.03 | 0.50 | 0.13 | <0.11 | ... | 1.03 | 0.05 |
| F469-2 | A1 | 2.06 | 0.13 | 0.26 | 0.14 | 0.68 | 0.16 | 0.09 | 0.04 | 0.81 | 0.07 |
| F469-2 | A2 | 1.63 | 0.20 | 0.42 | 0.20 | 0.60 | 0.21 | ... | ... | 0.88 | 0.13 |
| F469-2 | A3 | 1.30 | 0.07 | 0.38 | 0.03 | 0.59 | 0.14 | 0.14 | 0.02 | 1.79 | 0.08 |
| F469-2 | A4 | 1.68 | 0.23 | ... | ... | ... | ... | ... | ... | 0.73 | 0.14 |
| F530-3 | A1 | 1.20 | 0.07 | 0.38 | 0.04 | 0.40 | 0.04 | 0.14 | 0.03 | 1.39 | 0.08 |
| F530-3 | A2 | ... | ... | ... | ... | 0.23 | 0.10 | 0.10 | 0.10 | 1.68 | 0.23 |
| F558-1 | A1 | 4.97 | 0.55 | 0.76 | 0.13 | ... | ... | ... | ... | 0.58 | 0.11 |
| F558-1 | A2 | 5.32 | 0.97 | ... | ... | ... | ... | ... | ... | ... | ... |
| F561-1 | A1 | 2.71 | 0.96 | ... | ... | ... | ... | ... | ... | 0.66 | 0.38 |
| F561-1 | A2 | ... | ... | ... | ... | ... | ... | ... | ... | 2.81 | 2.94 |
| F561-1 | A3 | 3.13 | 0.25 | ... | ... | 0.51 | 0.07 | 0.06 | 0.05 | 0.79 | 0.09 |
| F563-V1 | A1 | 2.71 | 0.59 | ... | ... | ... | ... | ... | ... | 1.84 | 0.43 |
| F563-V1 | A2 | 0.86 | 0.24 | ... | ... | ... | ... | ... | ... | 0.57 | 0.20 |
| F563-V2 | A1 | 1.46 | 0.34 | 0.36 | 0.19 | 0.49 | 0.20 | ... | ... | 0.49 | 0.20 |
| F563-V2 | A2 | 3.08 | 0.48 | ... | ... | 0.27 | 0.13 | ... | ... | 0.92 | 0.19 |
| F563-V2 | A3 | 2.51 | 0.13 | 0.38 | 0.04 | 0.39 | 0.04 | 0.03 | 0.03 | 1.05 | 0.06 |
| F568-1 | A1 | 1.48 | 0.45 | ... | ... | ... | ... | ... | ... | 0.56 | 0.28 |
| F568-1 | A2 | 3.03 | 0.79 | ... | ... | ... | ... | ... | ... | 0.80 | 0.31 |
| F568-6 | S1A1 | 3.22 | 0.32 | ... | ... | ... | ... | ... | ... | 0.32 | 0.08 |
| F568-6 | S1A2 | 2.01 | 0.20 | ... | ... | ... | ... | ... | ... | 0.27 | 0.07 |
| F568-6 | S3A1 | 8.42 | 7.12 | ... | ... | ... | ... | ... | ... | ... | ... |
| F568-6 | S3A2 | 2.81 | 0.21 | 0.19 | 0.08 | 0.30 | 0.05 | ... | ... | 0.27 | 0.05 |
| F568-6 | S3A3 | 4.74 | 1.86 | ... | ... | ... | ... | ... | ... | 0.39 | 0.39 |
| F577-V1 | A1 | 3.78 | 0.99 | ... | ... | ... | ... | ... | ... | 0.72 | 0.30 |
| F577-V1 | A2 | 4.69 | 2.01 | ... | ... | ... | ... | ... | ... | 1.72 | 0.83 |
| F577-V1 | A3 | 4.32 | 0.28 | 0.21 | 0.08 | 0.47 | 0.06 | 0.01 | 0.04 | 1.00 | 0.08 |
| F577-V1 | A5 | 3.96 | 0.31 | 0.31 | 0.11 | 0.34 | 0.06 | <0.10 | ... | 1.45 | 0.13 |
| F583-1 | A2 | 4.15 | 1.40 | ... | ... | ... | ... | ... | ... | 0.89 | 0.43 |
| F583-1 | A3 | 1.35 | 0.08 | <0.52 | ... | 0.30 | 0.04 | 0.02 | 0.03 | 1.72 | 0.10 |
| F583-5 | A1 | 2.73 | 0.26 | ... | ... | 0.46 | 0.18 | ... | ... | 1.09 | 0.12 |
| F583-5 | A2 | 0.70 | 0.09 | ... | ... | ... | ... | ... | ... | 1.10 | 0.12 |
| F585-3 | S1A1 | 3.48 | 0.28 | <0.36 | ... | ... | ... | ... | ... | 1.34 | 0.13 |
| F585-3 | S1A2 | 3.03 | 0.17 | 0.48 | 0.25 | ... | ... | ... | ... | 1.38 | 0.09 |
| F585-3 | S1A3 | 3.87 | 0.64 | ... | ... | ... | ... | ... | ... | 0.84 | 0.20 |
| F585-3 | S2A1 | 4.27 | 2.47 | ... | ... | ... | ... | ... | ... | 0.75 | 0.69 |
| F585-3 | S2A2 | 2.75 | 0.37 | ... | ... | ... | ... | ... | ... | 1.07 | 0.18 |
| F585-3 | S2A3 | 3.82 | 0.32 | <0.71 | ... | ... | ... | ... | ... | 1.19 | 0.12 |
| F585-3 | S2A4 | 2.28 | 0.23 | ... | ... | ... | ... | ... | ... | 1.18 | 0.14 |
| F611-1 | A1 | 1.54 | 0.10 | ... | ... | 0.46 | 0.05 | <0.14 | ... | 1.14 | 0.08 |
| F611-1 | A2 | 1.25 | 0.13 | 0.42 | 0.18 | ... | ... | ... | ... | 0.76 | 0.10 |
| F746-1 | A1 | 3.36 | 0.12 | 0.79 | 0.03 | 0.38 | 0.02 | ... | ... | 1.30 | 0.05 |
| F746-1 | A2 | 5.97 | 0.31 | 1.44 | 0.09 | 0.31 | 0.04 | 0.08 | 0.05 | 1.08 | 0.07 |
| F746-1 | A3 | 3.87 | 0.55 | 1.11 | 0.20 | ... | ... | ... | ... | 1.15 | 0.21 |
| U 1230 | A1 | 2.13 | 0.36 | ... | ... | 0.15 | 0.13 | ... | ... | 0.87 | 0.19 |
| U 1230 | A2 | 2.64 | 0.19 | ... | ... | 0.57 | 0.27 | 0.12 | 0.05 | 0.86 | 0.08 |
| U 1230 | A3 | 2.01 | 0.09 | 0.13 | 0.07 | 0.46 | 0.06 | <0.26 | ... | 0.94 | 0.05 |
| U 1230 | A4 | 2.97 | 0.73 | ... | ... | 0.78 | 0.38 | ... | ... | 1.60 | 0.43 |
| U 1230 | A5 | 0.83 | 0.74 | 0.39 | 0.14 | 0.50 | 0.22 | ... | ... | 1.69 | 0.23 |
| U 5675 | A1 | 5.70 | 1.24 | ... | ... | ... | ... | ... | ... | 0.59 | 0.24 |
| U 5675 | A3 | 3.48 | 0.44 | ... | ... | ... | ... | ... | ... | 0.69 | 0.14 |
| U 5709 | A1 | 4.03 | 2.51 | ... | ... | ... | ... | ... | ... | 0.52 | 0.13 |
| U 5709 | A2 | 6.30 | 3.03 | ... | ... | ... | ... | ... | ... | 1.24 | 0.75 |
| U 5709 | A3 | 2.35 | 0.15 | ... | ... | ... | ... | ... | ... | 0.60 | 0.06 |
| U 6151 | S1A1 | 3.76 | 0.12 | 0.33 | 0.12 | 0.42 | 0.02 | ... | ... | 0.97 | 0.04 |
| U 6151 | S1A2 | 4.16 | 0.15 | 0.42 | 0.02 | 0.46 | 0.03 | 0.13 | 0.02 | 1.05 | 0.04 |
| U 6151 | S3A1 | 11.44 | 1.54 | ... | ... | ... | ... | ... | ... | 0.35 | 0.12 |
| U 6151 | S3A2 | 3.16 | 0.16 | ... | ... | ... | ... | ... | ... | 0.22 | 0.03 |
| U 6614 | A1 | 1.62 | 0.36 | ... | ... | ... | ... | ... | ... | 0.10 | 0.17 |
| U 6614 | A2 | 1.82 | 0.45 | ... | ... | ... | ... | ... | ... | 0.20 | 0.20 |
| U 6614 | A3 | 1.17 | 0.35 | ... | ... | ... | ... | ... | ... | 0.06 | 0.20 |
| U 9024 | S1A1 | 2.43 | 0.16 | 0.68 | 0.26 | 0.38 | 0.05 | <0.12 | ... | 1.59 | 0.11 |
| U 9024 | S1A2 | 3.27 | 0.27 | ... | ... | ... | ... | ... | ... | 0.59 | 0.08 |
| U 9024 | S1A3 | 2.69 | 0.61 | ... | ... | ... | ... | ... | ... | 0.45 | 0.22 |
| U 9024 | S2A1 | 3.00 | 0.18 | ... | ... | ... | ... | ... | ... | 1.06 | 0.07 |
| U12695 | S1A1 | 1.14 | 0.10 | 0.55 | 0.17 | 0.45 | 0.06 | 0.01 | 0.05 | 1.31 | 0.11 |
| U12695 | S1A2 | 2.01 | 0.07 | 0.43 | 0.12 | 0.35 | 0.02 | 0.03 | 0.02 | 1.37 | 0.05 |
| U12695 | S1A3 | 1.77 | 0.13 | 0.46 | 0.16 | 0.41 | 0.05 | ... | ... | 1.58 | 0.12 |
| U12695 | S2A1 | 0.73 | 0.03 | 0.44 | 0.12 | 0.44 | 0.02 | 0.12 | 0.02 | 1.78 | 0.06 |
| U12695 | S2A2 | 1.90 | 0.06 | 0.31 | 0.12 | 0.43 | 0.02 | 0.06 | 0.01 | 1.15 | 0.04 |
| U12695 | S2A3 | 1.90 | 0.10 | 0.26 | 0.13 | 0.43 | 0.04 | 0.06 | 0.03 | 1.52 | 0.08 |

TABLE 2 — *Continued*

| Galaxy | H II | [O III] λ5007 | $\sigma_\lambda$ | Hα λ6563 | $\sigma_\lambda$ | [N II] λ6583 | $\sigma_\lambda$ | [S II] λ6717 | $\sigma_\lambda$ | [S II] λ6731 | $\sigma_\lambda$ |
|---|---|---|---|---|---|---|---|---|---|---|---|
| F415-3 | A1 | 3.62 | 0.13 | 4.03 | 0.14 | 0.15 | 0.02 | 0.27 | 0.02 | 0.21 | 0.02 |
| F469-2 | A1 | 2.36 | 0.14 | 3.37 | 0.20 | 0.14 | 0.04 | 0.21 | 0.04 | 0.17 | 0.04 |
| F469-2 | A2 | 2.70 | 0.31 | 2.97 | 0.33 | <0.13 | ... | 0.18 | 0.09 | 0.28 | 0.09 |
| F469-2 | A3 | 5.81 | 0.24 | 3.22 | 0.14 | 0.07 | 0.02 | 0.15 | 0.02 | 0.14 | 0.02 |
| F469-2 | A4 | 1.24 | 0.19 | 4.72 | 0.58 | 0.28 | 0.10 | ... | ... | ... | ... |
| F530-3 | A1 | 3.88 | 0.19 | 4.06 | 0.20 | 0.19 | 0.03 | 0.44 | 0.04 | 0.27 | 0.03 |
| F530-3 | A2 | 3.65 | 0.46 | 3.14 | 0.40 | 0.15 | 0.10 | 0.20 | 0.10 | 0.044 | 0.10 |
| F558-1 | A1 | 1.29 | 0.17 | 3.27 | 0.37 | 0.59 | 0.11 | ... | ... | ... | ... |
| F558-1 | A2 | 0.30 | 0.17 | ... | ... | ... | ... | ... | ... | ... | ... |
| F561-1 | A1 | 2.44 | 0.87 | 2.94 | 1.03 | <0.22 | ... | 0.43 | 0.34 | 0.60 | 0.37 |
| F561-1 | A2 | 5.38 | 5.39 | 3.73 | 3.81 | ... | ... | ... | ... | ... | ... |
| F561-1 | A3 | 2.37 | 0.19 | 4.35 | 0.33 | <0.19 | ... | 0.66 | 0.08 | 0.53 | 0.07 |
| F563-V1 | A1 | 4.42 | 0.93 | 3.27 | 0.70 | <0.08 | ... | ... | ... | ... | ... |
| F563-V1 | A2 | 1.70 | 0.37 | 3.27 | 0.64 | <0.09 | ... | 0.31 | 0.18 | 0.23 | 0.17 |
| F563-V2 | A1 | 1.20 | 0.30 | 3.28 | 0.67 | <0.23 | ... | 0.96 | 0.26 | 0.63 | 0.22 |
| F563-V2 | A2 | 3.04 | 0.48 | 6.73 | 1.01 | <0.67 | ... | 1.21 | 0.23 | 1.02 | 0.21 |
| F563-V2 | A3 | 3.03 | 0.15 | 6.32 | 0.31 | 0.28 | 0.03 | 0.88 | 0.06 | 0.73 | 0.05 |
| F568-1 | A1 | 1.69 | 0.50 | ... | ... | ... | ... | ... | ... | ... | ... |
| F568-1 | A2 | 2.40 | 0.64 | ... | ... | ... | ... | ... | ... | ... | ... |
| F568-6 | S1A1 | 0.96 | 0.12 | 8.01 | 0.76 | ... | ... | ... | ... | ... | ... |
| F568-6 | S1A2 | 0.83 | 0.11 | 6.01 | 0.54 | ... | ... | ... | ... | ... | ... |
| F568-6 | S3A1 | ... | ... | 4.30 | 3.72 | 2.13 | 1.98 | 1.39 | 1.44 | 0.40 | 0.89 |
| F568-6 | S3A2 | 0.81 | 0.08 | 4.73 | 0.33 | 1.28 | 0.11 | 0.82 | 0.08 | 0.15 | 0.05 |
| F568-6 | S3A3 | 1.17 | 0.58 | 6.86 | 2.66 | 2.11 | 0.90 | 0.90 | 0.51 | 1.09 | 0.56 |
| F577-V1 | A1 | 2.06 | 0.58 | 4.88 | 1.26 | ... | ... | ... | ... | ... | ... |
| F577-V1 | A2 | 5.13 | 2.19 | 10.05 | 4.21 | <0.66 | ... | 2.55 | 1.15 | 3.30 | 1.44 |
| F577-V1 | A3 | 2.99 | 0.20 | 4.24 | 0.27 | 0.34 | 0.05 | 0.68 | 0.07 | 0.46 | 0.06 |
| F577-V1 | A5 | 4.21 | 0.33 | 4.17 | 0.32 | <0.37 | ... | ... | ... | ... | ... |
| F583-1 | A2 | 2.67 | 0.93 | 3.64 | 1.23 | ... | ... | ... | ... | ... | ... |
| F583-1 | A3 | 5.14 | 0.27 | 2.87 | 0.16 | ... | ... | ... | ... | ... | ... |
| F583-5 | A1 | 3.26 | 0.30 | ... | ... | ... | ... | ... | ... | ... | ... |
| F583-5 | A2 | 3.28 | 0.29 | ... | ... | ... | ... | ... | ... | ... | ... |
| F585-3 | S1A1 | 3.99 | 0.32 | 3.01 | 0.25 | <0.07 | ... | ... | ... | ... | ... |
| F585-3 | S1A2 | 4.13 | 0.23 | 3.01 | 0.17 | 0.35 | 0.04 | 0.43 | 0.04 | 0.34 | 0.04 |
| F585-3 | S1A3 | 3.26 | 0.54 | 3.01 | 0.50 | 0.41 | 0.15 | 0.39 | 0.15 | 0.56 | 0.17 |
| F585-3 | S2A1 | 2.51 | 1.52 | ... | ... | ... | ... | ... | ... | ... | ... |
| F585-3 | S2A2 | 3.43 | 0.45 | ... | ... | ... | ... | ... | ... | ... | ... |
| F585-3 | S2A3 | 4.15 | 0.35 | ... | ... | ... | ... | ... | ... | ... | ... |
| F585-3 | S2A4 | 3.43 | 0.34 | ... | ... | ... | ... | ... | ... | ... | ... |
| F611-1 | A1 | 3.45 | 0.20 | 3.91 | 0.22 | 0.24 | 0.04 | 0.36 | 0.04 | 0.17 | 0.04 |
| F611-1 | A2 | 2.68 | 0.24 | 3.36 | 0.30 | 0.11 | 0.06 | 0.17 | 0.06 | 0.12 | 0.06 |
| F746-1 | A1 | 4.06 | 0.14 | 3.71 | 0.13 | <0.11 | ... | ... | ... | ... | ... |
| F746-1 | A2 | 3.23 | 0.17 | 5.30 | 0.28 | 0.61 | 0.05 | 0.96 | 0.06 | 0.87 | 0.06 |
| F746-1 | A3 | 3.03 | 0.44 | ... | ... | ... | ... | ... | ... | ... | ... |
| U 1230 | A1 | 2.06 | 0.35 | 3.45 | 0.54 | <0.13 | ... | 0.45 | 0.15 | 0.30 | 0.14 |
| U 1230 | A2 | 2.75 | 0.20 | 3.40 | 0.24 | 0.28 | 0.05 | 0.42 | 0.06 | 0.27 | 0.05 |
| U 1230 | A3 | 2.51 | 0.11 | 3.35 | 0.14 | 0.26 | 0.02 | 0.33 | 0.03 | 0.28 | 0.03 |
| U 1230 | A4 | 3.92 | 0.94 | 5.52 | 1.30 | <0.17 | ... | 0.37 | 0.23 | 0.48 | 0.24 |
| U 1230 | A5 | 4.88 | 0.58 | 3.25 | 0.40 | 0.22 | 0.10 | 0.31 | 0.10 | 0.28 | 0.10 |
| U 5675 | A1 | 1.78 | 0.44 | ... | ... | ... | ... | ... | ... | ... | ... |
| U 5675 | A3 | 2.70 | 0.35 | ... | ... | ... | ... | ... | ... | ... | ... |
| U 5709 | A1 | 1.57 | 0.23 | 2.89 | 0.38 | 0.73 | 0.14 | ... | ... | ... | ... |
| U 5709 | A2 | 3.70 | 1.83 | 9.83 | 4.69 | 3.11 | 1.55 | ... | ... | ... | ... |
| U 5709 | A3 | 1.80 | 0.12 | 3.75 | 0.24 | 0.93 | 0.08 | ... | ... | ... | ... |
| U 6151 | S1A1 | 2.96 | 0.10 | 3.31 | 0.11 | <0.28 | ... | 0.56 | 0.03 | 0.36 | 0.02 |
| U 6151 | S1A2 | 3.09 | 0.11 | 5.18 | 0.18 | 0.18 | 0.02 | 0.94 | 0.04 | 0.55 | 0.03 |
| U 6151 | S3A1 | 1.05 | 0.19 | 2.86 | 0.41 | <0.11 | ... | 0.50 | 0.14 | 0.34 | 0.12 |
| U 6151 | S3A2 | 0.64 | 0.05 | 3.10 | 0.16 | <0.16 | ... | 0.43 | 0.04 | 0.21 | 0.03 |
| U 6614 | A1 | 0.31 | 0.18 | ... | ... | ... | ... | ... | ... | ... | ... |
| U 6614 | A2 | 0.61 | 0.24 | ... | ... | ... | ... | ... | ... | ... | ... |
| U 6614 | A3 | 0.19 | 0.21 | ... | ... | ... | ... | ... | ... | ... | ... |
| U 9024 | S1A1 | 4.75 | 0.29 | ... | ... | ... | ... | ... | ... | ... | ... |
| U 9024 | S1A2 | 2.25 | 0.19 | ... | ... | ... | ... | ... | ... | ... | ... |
| U 9024 | S1A3 | 1.34 | 0.35 | ... | ... | ... | ... | ... | ... | ... | ... |
| U 9024 | S2A1 | 3.16 | 0.18 | ... | ... | ... | ... | ... | ... | ... | ... |
| U12695 | S1A1 | 3.98 | 0.29 | 3.24 | 0.24 | <0.07 | ... | 0.12 | 0.05 | 0.09 | 0.05 |
| U12695 | S1A2 | 3.91 | 0.12 | 3.35 | 0.10 | 0.07 | 0.01 | 0.19 | 0.05 | 0.13 | 0.05 |
| U12695 | S1A3 | 4.33 | 0.29 | 2.65 | 0.18 | 0.04 | 0.03 | 0.08 | 0.04 | 0.06 | 0.04 |
| U12695 | S2A1 | 5.39 | 0.16 | ... | ... | ... | ... | ... | ... | ... | ... |
| U12695 | S2A2 | 3.27 | 0.10 | 3.26 | 0.10 | <0.09 | ... | 0.36 | 0.05 | 0.26 | 0.11 |
| U12695 | S2A3 | 4.40 | 0.22 | ... | ... | ... | ... | ... | ... | ... | ... |

TABLE 3
LSB ABUNDANCES

| Galaxy | H II | O/H | $\sigma_O$ | <U> | $\sigma_U$ |
|---|---|---|---|---|---|
| F415–3 | A1 | −3.94 | 0.11 | −2.65 | 0.11 |
| F469–2 | A1 | −4.17 | 0.11 | −2.73 | 0.12 |
| F469–2 | A2 | −4.29 | 0.21 | −2.53 | 0.17 |
| F469–2 | A3 | −3.98 | 0.11 | −2.07 | 0.11 |
| F469–2 | A4 | −4.24 | 0.25 | −3.02 | 0.15 |
| F530–3 | A1 | −4.14 | 0.09 | −2.29 | 0.11 |
| F561–1 | A3 | −3.86 | 0.23 | −3.03 | 0.13 |
| F563–V1 | A1 | −3.90 | 0.37 | −2.46 | 0.21 |
| F563–V1 | A2 | −4.67 | 0.25 | −2.49 | 0.22 |
| F563–V2 | A1 | −4.48 | 0.30 | −2.91 | 0.23 |
| F563–V2 | A2 | −3.60 | 0.49 | −3.01 | 0.24 |
| F563–V2 | A3 | −3.60 | 0.30 | −2.91 | 0.20 |
| F568–6 | S1A1 | −3.60 | 0.36 | −3.57 | 0.25 |
| F568–6 | S3A2 | −3.22 | 0.28 | −3.30 | 0.27 |
| F577–V1 | A3 | −3.60 | 0.20 | −3.00 | 0.21 |
| F577–V1 | A5 | −3.60 | 0.20 | −2.81 | 0.22 |
| F583–1 | A3 | −4.04 | 0.13 | −2.09 | 0.11 |
| F585–3 | S1A1 | −3.92 | 0.20 | −2.67 | 0.11 |
| F611–1 | A1 | −4.14 | 0.12 | −2.46 | 0.11 |
| F611–1 | A2 | −4.33 | 0.13 | −2.50 | 0.15 |
| F746–1 | A1 | −3.60 | 0.20 | −2.86 | 0.20 |
| F746–1 | A2 | −3.60 | 0.30 | −3.14 | 0.30 |
| F746–1 | A3 | −3.82 | 0.28 | −2.81 | 0.11 |
| U 1230 | A1 | −4.17 | 0.30 | −2.80 | 0.20 |
| U 1230 | A2 | −4.04 | 0.16 | −2.77 | 0.13 |
| U 1230 | A3 | −4.16 | 0.09 | −2.66 | 0.11 |
| U 1230 | A4 | −3.66 | 0.47 | −2.78 | 0.28 |
| U 5675 | A1 | −3.64 | 0.46 | −3.13 | 0.23 |
| U 5709 | A1 | −3.02 | 0.70 | −2.40 | 0.70 |
| U 5709 | A3 | −3.21 | 0.13 | −2.70 | 0.40 |
| U 6151 | S1A1 | −3.65 | 0.31 | −2.85 | 0.20 |
| U 6151 | S1A2 | −3.60 | 0.25 | −3.04 | 0.25 |
| U 6151 | S3A1 | −3.60 | 0.20 | −3.70 | 0.20 |
| U 6151 | S3A2 | −4.08 | 0.09 | −3.46 | 0.11 |
| U 9024 | S1A1 | −3.70 | 0.36 | −2.46 | 0.20 |
| U12695 | S1A1 | −4.20 | 0.12 | −2.20 | 0.15 |
| U12695 | S1A2 | −4.02 | 0.09 | −2.46 | 0.12 |
| U12695 | S1A3 | −4.02 | 0.15 | −2.30 | 0.13 |
| U12695 | S2A1 | −4.14 | 0.08 | −1.82 | 0.17 |
| U12695 | S2A2 | −4.11 | 0.08 | −2.50 | 0.11 |
| U12695 | S2A3 | −4.01 | 0.12 | −2.31 | 0.10 |

TABLE 4
Ambiguous Abundance Determinations

| Galaxy | H II | Lower Branch | | | | Upper Branch | | | |
|---|---|---|---|---|---|---|---|---|---|
| | | O/H | $\sigma_O$ | $<U>$ | $\sigma_U$ | O/H | $\sigma_O$ | $<U>$ | $\sigma_U$ |
| F558–1 | A1 | −3.92 | 0.30 | −3.35 | 0.14 | −3.28 | 0.22 | −3.17 | 0.22 |
| F568–1 | A2 | −4.10 | 0.31 | −2.85 | 0.22 | −3.27 | 0.22 | −2.62 | 0.21 |
| F583–5 | A1 | −4.08 | 0.20 | −2.65 | 0.10 | −3.31 | 0.16 | −2.44 | 0.16 |
| F583–5 | A2 | −4.41 | 0.18 | −2.09 | 0.17 | −3.16 | 0.18 | −1.64 | 0.21 |
| F585–3 | S1A2 | −4.28 | 0.26 | −2.06 | 0.12 | −3.24 | 0.14 | −1.72 | 0.17 |
| F585–3 | S1A3 | −3.98 | 0.26 | −2.85 | 0.11 | −3.35 | 0.19 | −2.70 | 0.16 |
| F585–3 | S2A2 | −4.00 | 0.21 | −2.64 | 0.11 | −3.36 | 0.18 | −2.45 | 0.16 |
| F585–3 | S2A3 | −3.95 | 0.24 | −2.70 | 0.10 | −3.38 | 0.21 | −2.52 | 0.15 |
| F585–3 | S2A4 | −4.10 | 0.23 | −2.53 | 0.17 | −3.32 | 0.21 | −2.31 | 0.24 |
| U 1230 | A5 | −4.19 | 0.30 | −1.98 | 0.30 | −3.30 | 0.30 | −1.69 | 0.30 |
| U 5675 | A3 | −3.99 | 0.20 | −2.85 | 0.12 | −3.33 | 0.16 | −2.68 | 0.17 |
| U 6614 | A1 | −4.51 | 0.23 | −3.59 | 0.27 | −2.91 | 0.19 | −3.31 | 0.40 |
| U 6614 | A2 | −4.48 | 0.27 | −3.30 | 0.22 | −2.99 | 0.20 | −2.94 | 0.29 |
| U 6614 | A3 | −4.65 | 0.27 | −3.63 | 0.38 | −2.84 | 0.20 | −3.25 | 0.75 |
| U 9024 | S1A2 | −4.03 | 0.21 | −2.94 | 0.17 | −3.29 | 0.22 | −2.74 | 0.20 |
| U 9024 | S1A3 | −4.21 | 0.32 | −3.10 | 0.13 | −3.16 | 0.23 | −2.83 | 0.19 |
| U 9024 | S2A1 | −3.95 | 0.21 | −2.69 | 0.17 | −3.39 | 0.18 | −2.51 | 0.20 |

# REFERENCES

Abia, C., & Rebolo, R. 1989, ApJ, 347, 186

Adler, D. S., Allen, R. J., & Lo, K. Y. 1991, ApJ, 382, 475

Alloin, D., Collin-Souffrin, S., Joly, M., & Vigroux, L. 1979, A&A, 78, 200 ApJ, 329, 572

Bertola, F., Burstein, D., & Buson, L. M. 1993, ApJ, 403, 573

Bothun, G. D., Schombert, J. M., Impey, C. D., & Schneider, S. E. 1990, ApJ, 360, 427

Bothun, G. D., Schombert, J. M., Impey, C. D., Sprayberry, D., & McGaugh, S. S. 1993, AJ, 106, 548

Burstein, D., & Heiles, C. 1984, ApJS, 54, 33

Caldwell, N., Kennicutt, R., Phillips, A. C., & Schommer, R. A. 1991, ApJ, 370, 526

Campbell, A. 1988, ApJ, 335, 644

Campbell, A., Terlevich, R., & Melnick, J. 1986, MNRAS, 223, 811

Diaz, A. I., Terlevich, E., Vilchez, J.M., Pagel, B. E. J., & Edmunds, M. G. 1991, MNRAS, 253, 245

Dopita, M. A., & Evans, I. N. 1986, ApJ, 307, 431

Freeman, K. C. 1970, ApJ, 160, 811

Garnett, D. R., & Shields, G. A. 1987, ApJ, 317, 82

Gilmore, G., & Wyse, R. F. G. 1991, ApJ, 367, L55

Henry, R. B. C., Pagel, B. E. J., Lasseter, D. F., & Chincarini, G. L. 1992, MNRAS, 258, 321

Izotov, Yu. I., Lipovetskii, V. A., Guseva, N. G., Kniazev, A., Yu., & Stepanian, J. A. 1990, Nature, 343, 238

Kennicutt, R. C. 1983, ApJ, 272, 54

Kennicutt, R. C. 1989, ApJ, 344, 685

Kunth, D., & Sargent, W. L. W. 1986, ApJ, 300, 496

Luppino, G. A. 1989, PASP, 101, 931

Maloney, P., & Black, J. H. 1988, ApJ, 325, 389

Matteucci, F. 1986, PASP, 98, 973

McCall, L. M., Rybski, P. M., & Shields, G. A. 1985, ApJS, 57, 1

McGaugh, S. S. 1991, ApJ, 380, 140

McGaugh, S. S. 1992, Ph.D. thesis, University of Michigan

McGaugh, S. S., & Bothun, G. D. 1993, AJ, in press

McGaugh, S. S., Bothun, G. D., & Schombert, J. M. 1993a, AJ, submitted

McGaugh, S. S., & Bothun, G. D., van der Hulst, J. M., & Schombert, J. M. 1993b, in preparation

Mo, H. J., McGaugh, S. S., & Bothun, G. D. 1993, MNRAS, in press

Nilson, P. 1973, *Uppsala General Catalog of Galaxies* (Uppsala, Sweden: Societatis Scientiarum Upsaliensis) (UGC)

Oey, M. S., & Kennicutt, R. C. 1993, ApJ, 411, 137

Olofsson, K. 1989, A&AS, 80, 317

Osterbrock, D. E. 1989, *Astrophysics of Gaseous Nebulae and Active Galactic Nuclei* (Mill Valley: University Science Books)

Pagel, B. E. J., Edmunds, M. G. 1981, ARA&A, 19, 77

Pagel, B. E. J., Edmunds, M. G., Blackwell, D. E., Chun, M. S., & Smith, G. 1979, MNRAS, 193, 219

Peimbert, M., Sarmiento, A., & Fierro, J. 1991, PASP, 103, 815

Phillipps, S., & Edmunds, M. G. 1991, MNRAS, 251, 84

Romanishin, W., Strom, K. M., & Strom, S. E. 1983, ApJS, 53, 105

Sage, L. J., Salzer, J. J., Loose, H.-H., & Henkel, C. 1992, A&A, 265, 19

Salzer, J. J., Alighieri, S. D. S., Matteucci, F., Giovanelli, R., and Haynes, M. P. 1991, AJ, 101, 1258

Savage, B. D., & Mathis, J. S. 1979, ARA&A, 17, 73

Schombert, J. M., & Bothun, G. D. 1988, AJ, 95, 1389

Schombert, J. M., Bothun, G. D., Impey, C. D., & Mundy, L. G. 1990, AJ, 100, 1523

Schombert, J. M., Bothun, G. D., Schneider, S. E., & McGaugh, S. S. 1992, AJ, 103, 1107

Scowen, P. A., Dufour, R. J., & Hester, J. J. 1992, AJ, 104, 92

Shields, G. A., Skillman, E. D., & Kennicutt, R. C. 1991, ApJ, 371, 82

Silva, D. R. 1991, Ph.D. thesis, University of Michigan

Skillman, E. D. 1989, ApJ, 347, 883

Skillman, E. D., Kennicutt, R. C., & Hodge, P. W. 1989, ApJ, 347, 875

Terlevich, R. 1985, in *Star Forming Dwarf Galaxies*, eds. D. Kunth, T. X. Thuan, & J. J. T. Van (Paris: Ed. Frontieres), 395

Thomsen, B., & Baum, W. A. 1987, ApJ, 315, 460

Torres-Peimbert, S., Peimbert, M., & Fierro, J. 1989, ApJ, 345, 186

Tyson, N. D., & Scalo, J. M. 1988, ApJ, 329, 618

van der Hulst, J. M., Skillman, E. D., Kennicutt, R. C., & Bothun, G. D. 1987, A&A, 177, 63

van der Hulst, J. M., Skillman, E. D., Smith, T. R., Bothun, G. D., McGaugh, S. S., & de Blok, W. J. G. 1993, AJ, 106, 548

Vila-Costas, M. B., & Edmunds, M. G. 1992, MNRAS, 259, 121

Vila-Costas, M. B., & Edmunds, M. G. 1993, preprint

von Hippel, T., & Bothun, G. 1990, AJ, 100, 403

Webster, B. L., Longmore, A. J., Hawarden, T. G., & Mebold, U. 1983, MNRAS, 205, 643

Webster, B. L., & Smith, M. G. 1983, MNRAS, 204, 743

# FIGURE CAPTIONS

**Figure 1.** Spectrum of H II region A3 in UGC 1230 with important nebular emission lines labeled. This spectrum is chosen for illustration because all the relevant lines are visible. Usually the weak lines [O III] $\lambda4363$, He I $\lambda5876$, and [N II] $\lambda6548$ are not detected, and [N II] $\lambda6583$ is generally weaker than it is here. Though [O III] $\lambda4363$ is detected, it is unphysically large. This can be caused by the shocks likely to be present within H II regions (Peimbert, Sarmiento, & Fierro 1991), but in this case is probably just a result of low signal to noise. Empirical methods of estimating the oxygen abundance utilizing the bright $R_{23}$ lines are superior to directly measuring the electron temperature with $\lambda4363$ until quite high signal to noise is obtained in this faint line.

**Figure 2.** LSB H II region data (solid squares) plotted against the model grid of McGaugh (1991). Solid lines refer to the upper branch ($\log(O/H) \geq -3.6$) of the $R_{23}$ calibration, while dashed lines define the lower branch. Vertical lines of constant abundance are separated by 0.1 in $\log(O/H)$. Several of these lines are labeled by the corresponding value of $\log(O/H)$. Horizontal lines of constant ionization parameter are separated by 1 in $\log<U>$ and range from $\log<U> = -4$ at the bottom to $\log<U> = -1$ at the top. This grid is for a set of H II regions ionized by a cluster IMF with $M_u = 60 M_\odot$. A larger $M_u$ would shift the grid to the right slightly, while a smaller $M_u$ would shift it to the left.

**Figure 3.** [N II] $\lambda6583$/[O II] $\lambda3727$ plotted against the abundance indicating line ratio $R_{23}$. The open triangles are data for HSB spirals from McCall et al. (1985). Solid squares are data for LSB galaxies. The LSB data begin where the HSB data trail off. The transition through the $R_{23}$ turnover region occurs around $\log([N\ II]/[O\ II]) \approx -1$, so the majority of LSB galaxies are on the lower branch.

*Inset:* The theoretically expected behavior of these line ratios. Above $\log([N\ II]/[O\ II]) \approx -1$, a narrow sequence is anticipated as neither abscissa nor ordinate is very sensitive to either the ionization parameter or the shape of the ionizing spectrum: this is a single parameter sequence in metallicity. Below this value, the scatter in the data is expected to increase due to the sensitivity of $R_{23}$ to $<U>$ on the lower branch. The exact path of models with constant $<U>$ depends on the N/O abundance ratio, which varies in a complicated way with O/H. Various possibilities are illustrated.

**Figure 4.** Nebular ionization parameters and oxygen abundances. This is the result of transforming Figure 2 from ($R_{23}$, $O_{32}$) to (O/H, $<U>$) coordinates. As in Figure 2, no correlation is evident, though an envelope of limiting $<U>$ which decreases with increasing O/H does appear to be present. Solid squares are data for which the branch assignment is certain; open squares represent data for which the branch determination is ambiguous but which can reasonably be assumed to be on the lower branch (see text). The stacking of points at $\log(O/H) = -3.6$ is an artefact of the $R_{23}$ method induced by the rapid change in the manifold of Figure 2 near this value of the abundance.

**Figure 5.** A histogram of the reddenings determined from the Balmer decrements and corrected for the galactic contribution. The hatched histogram is for LSB galaxies, while the open histogram is for HSB spirals (McCall et al. 1985).

**Figure 6.** The reddening towards LSB H II regions, corrected for the galactic contribution, plotted against the oxygen abundance. Though the errors are large, there is a clear trend for the dust content to increase with metallicity.

**Figure 7.** Histogram of oxygen abundances for LSB galaxies. The hatched histogram is for H II regions for which the branch assignment is certain. The open histogram represents the total sample when those H II regions with ambiguous abundance determinations which can reasonably be assumed to be on the lower branch are included. For reference, solar abundance and the abundance of the most metal poor galaxies known, I Zw 18 and SBS 0335 − 052, are marked with arrows.

The spike at $\log(O/H) = -3.6$ is *partly* an artefact of the method employed to determine the oxygen abundances.

**Figure 8.** The oxygen abundances of LSB galaxies plotted against their linear size as measured by the exponential scale length $\alpha$ (in kpc) of their disks. The stacking of points in the vertical direction is due to the observation of multiple H II regions within a single galaxy. Although the LSB disks cover the same range in size as Freeman disks (for the assumed short distance scale), and include some quite large disks, they are much more metal poor than a "typical" solar abundance ($\log(O/H) \approx -3.1$; near the top of the graph) HSB spiral galaxy. This holds regardless of size, and there is no obvious tendency for metallicity to increase with size, especially given the incomplete sampling of H II regions in the largest galaxy (Malin 2).

**Figure 9.** Oxygen abundance plotted against central surface brightness. Although selecting for low surface brightness has yielded low abundance objects, no direct correlation is apparent. While comparing local and global quantities is difficult (see text), this seems to indicate a complicated situation in which no single global property (such as size or surface brightness) determines the metallicity of a galaxy.

**Figure 10.** The distribution of LSB galaxies in the $O/H$–$M_B$ plane. Absolute magnitudes refer to the disk component only, the majority of these galaxies having little or no spheroidal component. Again, no correlation is apparent, and these LSB galaxies do not obviously follow the the mean relation (solid line) defined by dwarf irregulars (Skillman et al. 1989). Increasing the adopted short distance scale shifts the data to the right, generally further from the $1\sigma$ range of this relation (dashed lines). Excluding Malin 2, which is very much brighter than the objects for which the relation is defined, the LSB data make a scatter diagram.

**Figure 11.** The distribution of LSB galaxies in the oxygen abundance, dynamical mass plane. Masses (in $M_\odot$) are derived from H I line widths. Grossly similar to the O/H–

$M_B$ plane, no trend of metallicity increasing with mass is present, especially if the incompletely sampled giant Malin 2 is excluded.